\documentclass[12pt, a4paper]{elsarticle}

\usepackage[utf8]{inputenc}
\usepackage[OT1]{fontenc}
\usepackage{graphicx}
\usepackage[english]{babel}

\usepackage{amsmath}
\usepackage{amsfonts}
\usepackage{amssymb}
\usepackage{amsthm}

\usepackage[usenames,dvipsnames]{xcolor}
\usepackage{booktabs}
\usepackage{todonotes}
\usepackage{etoolbox}
\usepackage{url}

\usepackage[bookmarks]{hyperref}

\usepackage{mathtools}

\newtheorem{theorem}{Theorem}
\newtheorem{lemma}[theorem]{Lemma}

\newcommand{\bs}[1]{\boldsymbol{#1}}
\renewcommand{\vec}[1]{{\bs#1}}

\newcommand{\uz}{^{(0)}} 
\newcommand{\un}{^{(n)}} 

\newcommand{\ul}[1]{\underline{#1}}
\newcommand{\ol}[1]{\overline{#1}}

\newcommand{\Rsys}{R_\text{sys}}
\newcommand{\lRsys}{\ul{R}_\text{sys}}
\newcommand{\uRsys}{\ol{R}_\text{sys}}

\def\Tsys{T_\text{sys}}

\newcommand{\E}{\operatorname{E}}
\newcommand{\V}{\operatorname{Var}}

\newcommand{\indic}{\mathbb{I}}

\newcommand{\ber}{\operatorname{Bernoulli}} 
\newcommand{\bin}{\operatorname{Binomial}}
\newcommand{\be}{\operatorname{Beta}} 
\newcommand{\bebin}{\operatorname{Beta-Binomial}} 

\def\tmax{t_\text{max}}

\newcommand{\ptk}{p^k_t}

\def\yz{y\uz}
\def\yn{y\un}

\def\yl{\ul{y}}
\def\yu{\ol{y}}
\def\nl{\ul{n}}
\def\nu{\ol{n}}
\def\nktzo{\widetilde{n}\uz_{k,t}}
\def\pl{\ul{\psi}}
\def\pu{\ol{\psi}}
\def\el{\ul{\eta}}
\def\eu{\ol{\eta}}

\def\ykt{y_{k,t}}

\def\yktz{y\uz_{k,t}}
\def\yktn{y\un_{k,t}}

\def\yzl{\ul{y}\uz}
\def\yzu{\ol{y}\uz}
\def\ynl{\ul{y}\un}
\def\ynu{\ol{y}\un}

\def\yktzl{\ul{y}\uz_{k,t}}
\def\yktzu{\ol{y}\uz_{k,t}}
\def\yktnl{\ul{y}\un_{k,t}}
\def\yktnu{\ol{y}\un_{k,t}}

\newcommand{\ytz}[1]{y\uz_{#1,t}}
\newcommand{\ytzl}[1]{\ul{y}\uz_{#1,t}}
\newcommand{\ytzu}[1]{\ol{y}\uz_{#1,t}}

\newcommand{\ytnl}[1]{\ul{y}\un_{#1,t}}
\newcommand{\ytnu}[1]{\ol{y}\un_{#1,t}}

\def\nz{n\uz}
\def\nn{n\un}

\def\nktz{n\uz_{k,t}}
\def\nktn{n\un_{k,t}}

\def\nzl{\ul{n}\uz}
\def\nzu{\ol{n}\uz}

\def\nktzl{\ul{n}\uz_{k,t}}
\def\nktzu{\ol{n}\uz_{k,t}}

\newcommand{\ntzl}[1]{\ul{n}\uz_{#1,t}}
\newcommand{\ntzu}[1]{\ol{n}\uz_{#1,t}}

\def\MktZ{\mathcal{M}\uz_{k,t}}
\def\MktN{\mathcal{M}\un_{k,t}}

\def\PktZ{\Pi\uz_{k,t}}
\def\PktN{\Pi\un_{k,t}}
\newcommand{\PtZi}[1]{\Pi\uz_{#1,t}}
\newcommand{\PkZi}[1]{\Pi\uz_{k,#1}}

\newtoggle{td}
\newcommand{\td}[1]{%
  \iftoggle{td}{%
    \todo[inline]{#1}%
  }{}%
}

\allowdisplaybreaks

\toggletrue{td} 

\journal{journal}
\date{January 31, 2016}

\begin{document}


\begin{frontmatter}
\title{Bayesian Nonparametric System Reliability\\ using Sets of Priors}

\author[ein]{Gero Walter}
\ead{g.m.walter@tue.nl}
\author[oxf]{Louis J.M. Aslett}
\ead{louis.aslett@stats.ox.ac.uk}
\author[dur]{Frank P.A. Coolen}
\ead{frank.coolen@durham.ac.uk}

\address[ein]{School of Industrial Engineering, Eindhoven University of Technology, Eindhoven, NL}
\address[oxf]{Department of Statistics, University of Oxford, Oxford, UK}
\address[dur]{Department of Mathematical Sciences, Durham University, Durham, UK}

\begin{abstract}
An imprecise Bayesian nonparametric approach
to system reliability with multiple types of components
is developed.
This allows modelling partial or imperfect prior knowledge on component failure distributions
in a flexible way through bounds on the functioning probability.
Given component level test data these bounds are propagated 
to bounds on the posterior predictive distribution for the functioning probability of
a new system containing components exchangeable with those 
used in testing.  The method further enables identification 
of prior-data conflict at the system level based on component level test data.
New results on first-order stochastic dominance for the Beta-Binomial distribution
make the technique computationally tractable.
Our methodological contributions can be immediately used in
applications by reliability practitioners
as we provide easy to use software tools.
\end{abstract}

\end{frontmatter}


\section{Introduction}

System reliability analysis is concerned with
estimating the lifetime $\Tsys$ of complex systems.
Usually, the goal is to determine the system reliability function $\Rsys(t) = P(\Tsys > t)$
based on the lifetime distributions of system components.

A critique of the methodological approach to a reliability 
analysis may often encompass a few common concerns.
First, in a parametric 
setting, there may be no particularly strong reason to
believe that the small part of component model
space covered by a particular probability distribution 
necessarily contains the `true' component lifetime distribution.
Further, Bayesian methods may be invoked in order to 
incorporate expert opinion or other knowledge which falls
outside the specific testing data under consideration.  The 
classic concern here is in whether one can truly express 
ones beliefs with the exactness a prior distribution requires.  Finally,
it would be valuable in application to have a means of identifying when the
prior choice is having a strong effect and when not.  Any method 
hoping to address these concerns must do so whilst enabling 
realistic system models (with heterogeneous component types)
and remain computationally tractable.

Herein, we make steps toward addressing these concerns by developing
a nonparametric method which utilises imprecise probability \cite{itip,1991:walley}
to model more vague or imperfect prior beliefs using
upper and lower probabilities.  This overcomes the concern
about component lifetimes being outside a particular 
parametric family, uses a more flexible prior modelling 
framework and leads to an easy method of detecting 
conflicts between prior assumptions and observed failure times in test data.
In the general context of Bayesian methods,
this phenomenon is known as \emph{prior-data conflict},
see, e.g., \cite{2006:evans} or \cite{2015:bickel}.

Furthermore, the method is based on the survival signature \cite{2012:survsign},
a recent development which naturally accommodates 
heterogeneous component types laid out in any arbitrary manner.
By separating the (time-invariant) system structure from the time-dependent failure probabilities of components,
it allows straightforward and efficient computation of $\Rsys(t)$.

Our imprecise probability approach provides bounds for $\Rsys(t)$
by computing, for each $t$ in an arbitrarily fine grid of time points ${\cal T}$,
the posterior predictive probability interval for the event $\Tsys > t$. 
Assuming the number of functioning components for each type and time $t$ as binomially distributed, 
the intervals are derived from an imprecise Bayesian model using sets of conjugate Beta priors
which allow to specify weak or partial prior information in an intuitive way.
The width of the resulting posterior predictive probability intervals
reflects the precision of the corresponding probability statements:
a short range indicates that the system functioning probability can be quantified quite precisely,
while a large range will indicate that our (probabilistic) knowledge is indeterminate.
In particular, prior-data conflict leads to more cautious probability statements:
When there is not enough data to overrule the prior,
it is unclear whether to put more trust to prior assumptions or to the observations,
and posterior inferences clearly reflect this state of uncertainty by larger ranges.

While the use of imprecise probability methods can often lead to tractability issues,
new results on first-order stochastic dominance for the Beta-Binomial distribution
keep the need for numerical optimization in our model to a minimum.

In Section \ref{sec:survsign} we review the survival 
signature and in Section \ref{sec:nonparamapproach} we 
review the nonparametric approach to Bayesian reliability analysis
upon which our work builds 
\cite{2015:bayessurvsign}.  Section \ref{sec:reparam} details
the reparameterisation of that approach which enables the natural
formulation of the system reliability bounds, leading to
nice closed form results in some later theory.
Section \ref{sec:setsofbetapriors} lays the ground work to incorporate imprecise probability,
culminating in the main results and contributions of this 
work, detailed in Section \ref{sec:setsofrel}.  Section 
\ref{sec:examples} provides details on the software contributions
of this work and shows two worked examples
demonstrating the practicality and usefulness of the method.

\section{Survival Signature}
\label{sec:survsign}

In the mathematical theory of reliability, the main focus is on the functioning of a system given the functioning, or not, 
of its components and the structure of the system. The mathematical concept which is central to this theory is the 
\emph{structure function} \citep{BP75}. For a system with $m$ components, let state vector 
$\underline{x} = (x_1,x_2,\ldots,x_m) \in \{0,1\}^m$, with $x_i=1$ if the $i$th component functions 
and $x_i=0$ if not. The labelling of the components is arbitrary but must be fixed to define $\underline{x}$. 
The structure function $\phi : \{0,1\}^m \rightarrow \{0,1\}$, defined for all possible $\underline{x}$, takes 
the value 1 if the system functions and 0 if the system does not function for state vector $\underline{x}$. 
Most practical systems are coherent, which means that $\phi(\underline{x})$ 
is non-decreasing in any of the components of $\underline{x}$, so system functioning cannot be improved by worse performance 
of one or more of its components. The assumption of coherent systems is also convenient from the perspective of uncertainty
quantification for system reliability. It is further logical to assume that $\phi(\underline{0})=0$ and $\phi(\underline{1})=1$, 
so the system fails if all its components fail and it functions if all its components function. 

For larger systems, working with the full structure function may be complicated, and one may particularly
only need a summary of the structure function in case the system has exchangeable components of one or more
types. We use the term `exchangeable components' to indicate that the failure times of the components in the system
are exchangeable \citep{DF74}. \citet{2012:survsign} introduced such a summary,
called the \emph{survival signature}, 
to facilitate reliability analyses for systems with multiple types of components. In case of just a single type of components, 
the survival signature is closely related to the system signature \citep{Sa07}, which is well-established and the topic of many
research papers during the last decade. However, generalization of the signature to systems with
multiple types of components is extremely complicated (as it involves ordering order statistics of different
distributions), so much so that it cannot be applied to most practical systems. In addition to the 
possible use for such systems, where the benefit only occurs if there are multiple components of the 
same types, the survival signature is arguably also easier to interpret than the signature. 

Consider a system with $K\ge 1$ types of components, with $m_k$ components of type $k \in \{1,\ldots,K\}$ and 
$\sum_{k=1}^K m_k = m$. Assume that the random failure times of components of the same type are exchangeable \citep{DF74}.
Due to the arbitrary ordering of the components in the state vector, components of the same type can be grouped together, 
leading to a state vector that can be written as 
$\underline{x} = (\underline{x}^1,\underline{x}^2,\ldots,\underline{x}^K)$, with 
$\underline{x}^k = (x^k_1,x^k_2,\ldots,x^k_{m_k})$ the sub-vector representing the states of the components of type $k$. 

The \emph{survival signature} for such a system, denoted by $\Phi(l_1,\ldots,l_K)$, with $l_k=0,1,\ldots,m_k$ 
for $k=1,\ldots,K$, is defined as the probability for the event that the system functions given that \emph{precisely} $l_k$ of its 
$m_k$ components of type $k$ function, for each $k\in \{1,\ldots,K\}$ \citep{2012:survsign}.
Essentially, this creates a $K$-dimensional partition for the event $\Tsys > t$, such that $\Rsys(t) = P(\Tsys > t)$
can be calculated using the law of total probability:
\begin{align}
\label{eq:rsyswithsurvsign}
P(\Tsys > t)
 &= \sum_{l_1=0}^{m_1} \cdots \sum_{l_K=0}^{m_K} P(\Tsys > t \mid C^1_t = l_1,\ldots, C^K_t = l_K) \nonumber\\
 &  \hspace*{24ex}                        \times P\Big( \bigcap_{k=1}^K \{ C^k_t = l_k\} \Big) \nonumber\\
 &= \sum_{l_1=0}^{m_1} \cdots \sum_{l_K=0}^{m_K} \Phi(l_1, \ldots, l_K)
                                                 P\Big( \bigcap_{k=1}^K \{ C^k_t = l_k\} \Big) \,,
\end{align}
where $C^k_t \in \{0, 1, \ldots, m_k\}$ denotes
the random number of components of type $k$ functioning at time $t$. 

For calculating the survival signature based on the structure function, observe that
there are $\binom{m_k}{l_k}$ state vectors $\underline{x}^k$ with $\sum_{i=1}^{m_k} x^k_i = l_k$. Let $S^k_{l_k}$ 
denote the set of these state vectors for components of type $k$ and let $S_{l_1,\ldots,l_K}$ denote the set of 
all state vectors for the whole system for which $\sum_{i=1}^{m_k} x^k_i = l_k$, $k=1,\ldots,K$. Due to the 
exchangeability assumption for the failure times of the $m_k$ components of type $k$, all the state vectors 
$\underline{x}^k \in S^k_{l_k}$ are equally likely to occur, hence \citep{2012:survsign}
\begin{align}
\label{eq:surv-sig}
\Phi(l_1,\ldots,l_K)
 &= \left[ \prod_{k=1}^K \binom{m_k}{l_k}^{-1} \right] \times \sum_{\underline{x} \in S_{l_1,\ldots,l_K}} \phi(\underline{x})\,.
\end{align}
%
It should be emphasized that when using the survival signature,
there are no restrictions on dependence of the failure times of components of different types,
as the probability $P(\bigcap_{k=1}^K \{C^k_t = l_k\})$ can take any form of dependence into account,
for example one can include common-cause failures quite straightforwardly into this approach \cite{CCM15}. 
However, there is a substantial simplification
if one can assume that the failure times of components of different types are independent,
and even more so if one can assume that the failure times of components of type $k$ 
are conditionally independent and identically distributed with CDF $F_k(t)$.
With these assumptions, we get
\begin{align*}
\Rsys(t) &= \sum_{l_1=0}^{m_1} \cdots \sum_{l_K=0}^{m_K} \left[ \Phi(l_1,\ldots,l_K)
            \prod_{k=1}^K \left( \binom{m_k}{l_k} [F_k(t)]^{m_k-l_k} [1-F_k(t)]^{l_k} \right) \right]\,.
\end{align*}
We will employ both assumptions in this paper,
leading to $C^k_t$ having a Beta-Binomial distribution,
giving us a closed form expression for $P(C^k_t = l_k)$ for all $t$, $k$, and $l_k$.

The main advantage of the survival signature, in line with this property of the signature for systems with a single type of 
components \citep{Sa07}, is that the information about the system structure is fully 
separated from the information about functioning of the components, which simplifies related statistical inference as well as
considerations of optimal system design. In particular for study of system reliability over time, with the structure of the system, 
and hence the survival signature, not changing, this separation also enables relatively straightforward statistical inferences. 

There are several relatively straightforward generalizations of the use of the survival signature.
The probabilities for the numbers of functioning components can be generalized to lower and upper probabilities,
as e.g.\ done by \citet{CCMA14} within the nonparametric predictive inference (NPI) framework of statistics \citep{Co11},
where lower and upper probabilities for the events $C_k = l_k$
are inferred from test data on components of the same types as those in the system.
This is an approach that is also followed in the current paper, but with the use of generalized Bayesian inference instead of NPI.
Like \citet{CCMA14}, we will utilize the monotonicity of the survival signature for coherent systems
to simplify computations.

\section{Nonparametric Bayesian Approach for Component Reliability}
\label{sec:nonparamapproach}

Let us denote the random failure time of component number $i$ of type $k$ by $T^k_i$, $i = 1, \ldots, m_k$.
The failure time distribution can be written in terms of the cdf $F^k(t) = P(T^k_i \le t)$,
or in terms of the reliability function $R^k(t) = P(T^k_i > t) = 1 - F^k(t)$,
also known as the survival function.
For a nonparametric description of $R^k(t)$,
we consider a set of time points $t$, $t \in {\cal T} = \{t_1, \ldots, \tmax\}$.

At each time point $t$, the operational state of a single component of type $k$
is Bernoulli distributed (functioning: 1, failed: 0) with parameter $\ptk$, so that
\begin{align*}
P\big(\indic(T^k_i > t) = 1\big) &= \ptk\,, \\
P\big(\indic(T^k_i > t) = 0\big) &= 1 - \ptk\,,
\end{align*}
That is, $\indic(T^k_i > t) \sim \ber(\ptk)$, $i = 1, \ldots, m_k$, $t \in {\cal T}$.

The set of probabilities $\{ \ptk, t \in {\cal T}\}$
defines a discrete failure time distribution for components of type $k$ through
\begin{align*}
R^k(t_j) &= P(T^k > t_j) = p^k_{t_j},\ t_j = t_1, \ldots, \tmax\,.
\end{align*}
We can also express this failure time distribution through the probability mass function (pmf) and discrete hazard function,
\begin{align*}
f^k(t_j) &= P\big(T^k \in (t_j,t_{j+1}]\big) = p^k_{t_j} - p^k_{t_{j+1}}\,,\\ 
h^k(t_j) &= P\big(T^k \in (t_j,t_{j+1}]\mid T^k > t_j\big) = \frac{f^k(t_j)}{R^k(t_j)}\,. 
\end{align*}
The time grid $\cal T$ can be chosen to be appropriately dense for the application at hand,
with the natural extension between grid points by taking $R^k(\cdot)$ to be the right continuous step function induced by the grid values,
$R^k(t) = p^k_{t_j}, t \in [t_j, t_{j+1})$,
or by taking $p^k_{t_j}$ and $p^k_{t_{j+1}}$ as upper and lower bounds for $R^k(t)$, $t \in [t_j, t_{j+1})$.

The independence assumption for components of the same type immediately implies that 
the number of functioning components of type $k$ in the system
is binomially distributed, $C^k_t = \sum_{i=1}^{m_k} \indic(T^k_i > t) \sim \bin(\ptk, m_k)$.

The $\ptk$'s can, in theory, be directly chosen to arbitrarily closely approximate any valid lifetime pdf on
$[0,\infty)$, for example matching a bathtub curve for the corresponding hazard rate $h^k(t_j)$.
Naturally, $p^k_{t_j} \ge p^k_{t_{j+1}}$ should hold (assuming no repair).
However, such direct specification is non-trivial, neglects any inherent uncertainty in the particular choice, and cannot be 
easily combined with test data.
To account for the uncertainty, one can express knowledge about $\ptk$ through a prior distribution.
A convenient and natural choice is $\ptk \sim \be(\alpha^k_t, \beta^k_t)$, particularly because in a Bayesian inferential
setting this is the conjugate prior which leads to a Beta posterior.

Let the lifetime test data collected on component $k$ be $\vec{t}^k = (t^k_1, \ldots, t^k_{n_k})$.
At each fixed time $t \in {\cal T}$, this corresponds to an observation from the Binomial model described above,
$s^k_t = \sum_{i=1}^{n_k} \indic(t^k_i > t)$.
The posterior is then $\ptk \mid s^k_t \sim \be(\alpha^k_t + s^k_t, \beta^k_t + n_k - s^k_t)$.
The combination of a Binomial observation model with a Beta prior is often called Beta-Binomial model.

The posterior predictive distribution for the number of components surviving at time $t$
in a new system, based on the lifetime data and the prior information,
is then given by a so-called Beta-Binomial distribution,
$C^k_t \mid s^k_t \sim \bebin(m_k, \alpha^k_t + s^k_t, \beta^k_t + n_k - s^k_t)$.
That is, we have
\begin{align*}
P(C^k_t = l_k \mid s^k_t) &= {m_k \choose l_k} \frac{B(l_k + \alpha^k_t + s^k_t, m_k - l_k + \beta^k_t + n_k - s^k_t)}
                                                    {B(\alpha^k_t + s^k_t, \beta^k_t + n_k - s^k_t)} \,,
\end{align*}
where $B(\cdot, \cdot)$ is the Beta function.
This posterior predictive distribution is essentially a Binomial distribution
where the functioning probability $\ptk$ takes values in $[0,1]$
with the probability given by the posterior on $\ptk$.

Consequently, in order to calculate the system reliability according to \eqref{eq:rsyswithsurvsign},
for each component type $k$
we need lifetime data $\vec{t}^k$,
and have to choose $2 \times |{\cal T}|$ parameters
to specify the prior distribution for the discrete survival function of $T^k_i$.
In total, values for $2 \times |{\cal T}| \times K$ parameters must be chosen.

\section{Reparametrisation of the Beta Distribution}
\label{sec:reparam}

The parametrisation of the Beta distribution used above is common,
and allows $\alpha^k_t$ and $\beta^k_t$ to be interpreted as
hypothetical numbers of functioning and failed components of type $k$ at time $t$, respectively.
However, when we generalise to sets of priors in the sequel,
it is useful to consider a different parametrisation.

For clarity of presentation we will temporarily drop the super- and subscript $k$ and $t$ indices for component type and time.
Instead of $\alpha$ and $\beta$, we consider the parameters $\nz \in [0, \infty)$ and $\yz \in [0,1]$, where
\begin{align}
\nz &= \alpha + \beta &
&\text{and} &
\yz &= \frac{\alpha}{\alpha + \beta} \,,
\label{eq:reparam}
\end{align}
or equivalently, $\alpha = \nz\yz$ and $\beta = \nz(1-\yz)$.
The upper index ${}\uz$ is used to identify these as prior parameter values,
in contrast to their posterior values $\nn$ and $\yn$
obtained after observing $n$ failure times (see below).
$\nz$ and $\yz$ are sometimes called \emph{canonical} parameters,
identified from rewriting the density in canonical form;
see for example \cite[pp.~202 and 272f]{2000:bernardosmith}, or \cite[\S 1.2.3.1]{2013:diss-gw}.
This canonical form gives a common structure to all conjugacy results in exponential families.

From the properties of the Beta distribution,
it follows that $\yz = \E[p]$ is the prior expectation for the functioning probability $p$,
and that larger $\nz$ values lead to greater concentration of probability measure around $\yz$,
since $\V(p) = \frac{\yz (1-\yz)}{\nz + 1}$.
Consequently, $\nz$ represents the prior strength and moreover can be directly interpreted as a pseudocount, as will become clear.
Indeed, consider the posterior given that $s$ out of $n$ components function:
by conjugacy
$p \mid s$ is Beta distributed with updated parameters
\begin{align}
\nn &= \nz + n\,, &
\yn &= \frac{\nz}{\nz + n} \cdot \yz + \frac{n}{\nz + n} \cdot \frac{s}{n}\,.
\label{eq:nyupdate}
\end{align}
Thus, after observing that $s$ out of $n$ components function (at time $t$),
the posterior mean $\yn$ for $p$ is a weighted average of
the prior mean $\yz$ and $s/n$ (the fraction of functioning components in the data),
with weights proportional to $\nz$ and $n$, respectively.
Therefore $\nz$ takes on the same role for the prior mean $\yz$
as the sample size $n$ does for the observed mean $s/n$,
leading to the notion of it being a pseudocount.

Reintroducing time and component type indices, the posterior predictive Beta-Binomial probability mass function (pmf) can be written in terms of the updated parameters as
\begin{align}
P(C^k_t = l_k \mid s^k_t) &= {m_k \choose l_k} \frac{B(l_k + \nn_{k,t}\yn_{k,t}, m_k - l_k + \nn_{k,t}(1-\yn_{k,t}))}
                                                    {B(\nn_{k,t}\yn_{k,t}, \nn_{k,t}(1-\yn_{k,t}))} \,,
\label{eq:postpredCny}
\end{align}
with the corresponding cumulative mass function (cmf) given by
\begin{align}
F_{C^k_t\mid s^k_t}(l_k) &= P(C^k_t \le l_k \mid s^k_t) = \sum_{j_k=0}^{l_k} P(C^k_t = j_k \mid s^k_t)\,.
\label{eq:postpredCnycmf}
\end{align}

The parameterisation in terms of prior mean and prior strength (or pseudocount)
makes clear that in this conjugate setting,
learning from data corresponds to averaging between prior and data.
This form is attractive not only because it enhances the interpretability of the model and prior specification,
but crucially it also makes clear what should be a serious concern in any Bayesian analysis:
when observed data differ greatly from what is expressed in the prior,
this conflict is simply averaged out
and is not reflected in the posterior or posterior predictive distributions.

\begin{figure}
\includegraphics[width=\textwidth]{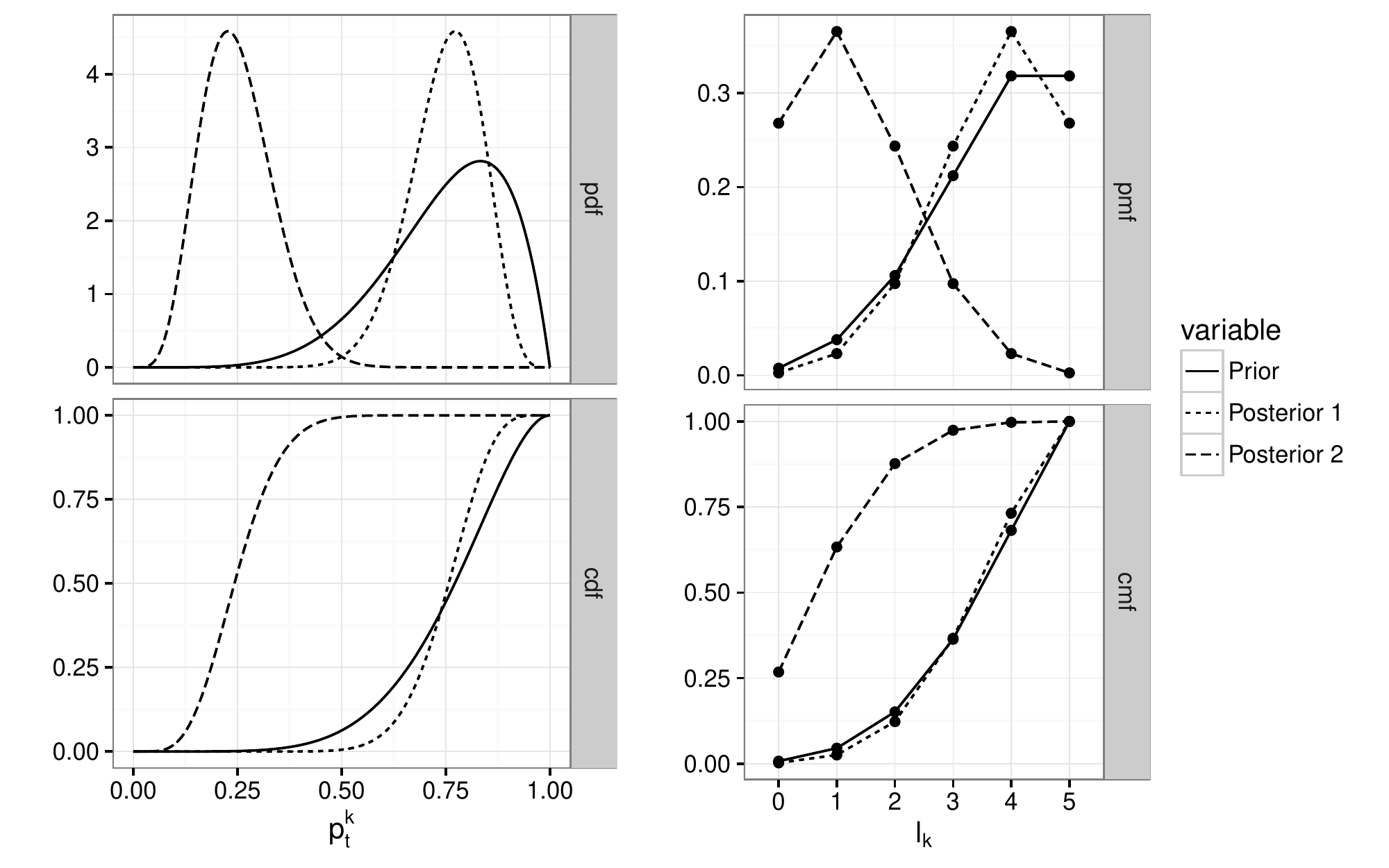}
\caption{Beta densities (top left) and cdfs (bottom left),
with the corresponding Beta-binomial predictive probability mass functions (top right) and cumulative mass functions (bottom right),
for a prior with $\nktz = 8, \yktz = 0.75$,
and posteriors based on $n^k_t=16$ observations with $s^k_t=12$ (Posterior~1) and $s^k_t=0$ (Posterior~2), respectively.
Data for Posterior~1 confirm prior assumptions,
while data for Posterior~2 are in conflict with the prior.
However, this conflict is averaged out,
and Posterior 1 and Posterior 2 have the same spread, both in the posterior pdf/cdf and the posterior predictive pmf/cmf,
such that Posterior 2 gives a false sense of certainty despite the massive conflict between prior and data.}
\label{fig:singleprior-pdc}
\end{figure}

As a simple example, imagine that we expect $\ptk$ to be about $0.75$ for a certain $k$ and $t$,
so we choose $\yktz = 0.75$,
and that we value this choice of mean functioning probability with $\nktz = 8$,
i.e., equivalently to having seen $8$ observations with a mean $0.75$.
If we observe $n_k = 16$ components of type $k$ in the test data and $s^k_t = 12$ function at time $t$,
then $s^k_t/n_k = 0.75$ as we expect,
so that the updated parameters are $\nktn = 24, \yktn=0.75$.
However, in contrast, unexpectedly observing that no component functions at time $t$ instead
leads to parameters $\nktn = 24, \yktn=0.25$.
The prior and the posteriors based on these two scenarios
are depicted in the left panels of Figure~\ref{fig:singleprior-pdc},
along with their corresponding predictive Beta-binomial pmf and cmf
for the case $m_k = 5$ (right panels).

Due to symmetry, we see that both posteriors have the same variance,
although arising from two fundamentally different scenarios.
Posterior 1 is based on data exactly according to prior expectations;
the increase in confidence on $\ptk \approx 0.75$
is reflected in a more concentrated posterior density,
and the posterior predictive is changed only slightly.
However, it may be cause for concern to see the same degree of confidence in Posterior 2,
which is based on data that is in sharp conflict with prior expectations.
Posterior 2 places most probability weight around $0.25$,
averaging between prior expectation and data,
with the same variance as Posterior 1.
Accordingly, rather than conveying the conflict between observed
and expected functioning probabilities with increased variance,
Posterior 2 instead gives a false sense of certainty.

To enable diagnosis of when this undesirable behaviour occurs,
we propose to use an imprecise probability approach
based on sets of Beta priors, described in the following section.

\section{Sets of Beta Priors}
\label{sec:setsofbetapriors}

As was shown by \citet{2009:WalterAugustin}, 
we can have both tractability and meaningful reaction to prior-data conflict
by using sets of priors $\MktZ$ produced by parameter sets $\PktZ = [\nktzl, \nktzu] \times [\yktzl, \yktzu]$
(a detailed discussion of different choices for $\PktZ$ is given in \citet[\S 3.1]{2013:diss-gw}.)
In our model, each prior parameter pair $(\nktz, \yktz) \in \PktZ$
corresponds to a Beta prior, thus $\MktZ$ is a set of Beta priors.
The set of posteriors $\MktN$ is obtained by updating each prior in $\MktZ$ according to Bayes' Rule.
This element-by-element updating can be rigorously justified
as ensuring coherence \citep[\S 2.5]{1991:walley}, and was termed ``Generalized Bayes' Rule'' by \citet[\S 6.4]{1991:walley}.
Due to conjugacy, $\MktN$ is a set of Beta distributions with parameters $(\nktn, \yktn)$,
obtained by updating $(\nktz, \yktz) \in \PktZ$ according to \eqref{eq:nyupdate},
leading to the set of updated parameters
\begin{align}
\PktN &= \Big\{ (\nktn, \yktn) \mid (\nktz, \yktz) \in \PktZ = [\nktzl, \nktzu] \times [\yktzl, \yktzu] \Big\}\,.
\label{eq:paramsets}
\end{align}

\begin{figure}
\includegraphics[width=\textwidth]{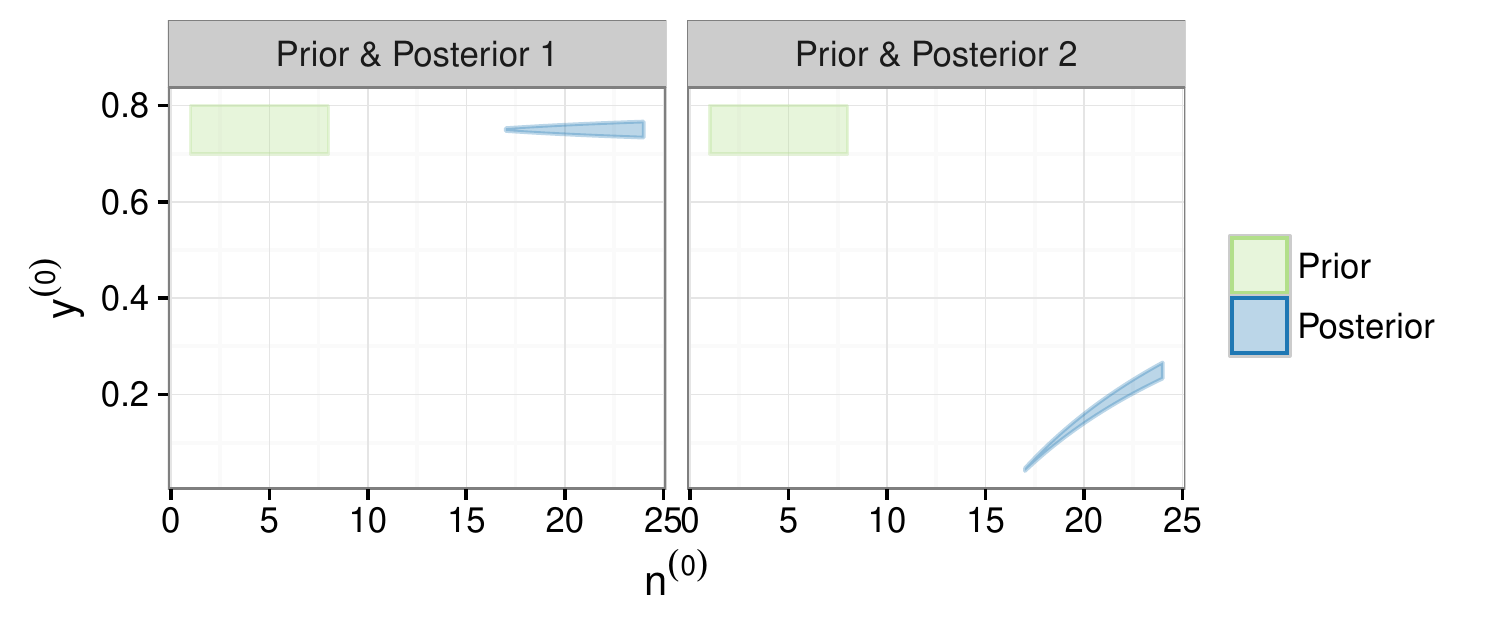}
\caption{Prior parameter set $\PktZ = [1,8] \times [0.7,0.8]$ and posterior parameter set $\PktN$
for data $s^k_t/n_k = 12/16$ (Posterior 1, left) and $s^k_t/n_k = 0/16$ (Posterior 2, right).
For no-conflict data ($s^k_t/n_k \in [\yktzl,\yktzu]$), $\PktN$ has the `spotlight' shape (left);
in case of prior-data conflict ($s^k_t/n_k \not\in [\yktzl,\yktzu]$), $\PktN$ has the `banana' shape (right),
leading to a large degree of imprecision in the $\yktn$ dimension of $\PktN$,
thus reflecting increased uncertainty about the functioning probability $p^k_t$
due to the conflict between prior assumptions and observed data.}
\label{fig:paramsets}
\end{figure}

Examples for parameter sets $\PktZ$ and $\PktN$ as in \eqref{eq:paramsets} are depicted in Figure~\ref{fig:paramsets}.
Such rectangular prior parameter sets $\PktZ$ have been shown
to balance desirable model properties and ease of elicitation 
(see \citet[pp.~123f]{2013:diss-gw} or \citet{Troffaes2013a-short}).
For each component type $k$ and time point $t$,
one need only specify the four parameters $\nktzl, \nktzu, \yktzl, \yktzu$
(so in total $4 \times |{\cal T}|$ parameters are needed to define the set of prior distributions
on the survival function of each component).

A desirable inference property arising from this setup is that
the posterior parameter set $\PktN$ is not rectangular in the way that the prior parameter set is.
Indeed, the shape of $\PktN$ depends on the presence or absence of prior-data conflict,
which is naturally operationalised as $s^k_t/n_k \not\in [\yktzl, \yktzu]$:
that is, prior-data conflict is defined to occur when, at time $t$, the observed fraction of functioning components
is outside its \emph{a priori} expected range.

First, in the absence of prior-data conflict, 
$\PktN$ shrinks in the $\ykt$ dimension;
how much it shrinks depending on $\nktz \in [\nktzl, \nktzu]$,
leading to the so-called spotlight shape depicted in Figure~\ref{fig:paramsets} (left).
Since $\yktn$ gives the posterior expectation for the functioning probability $p_t^k$,
shorter $\yktn$ intervals mean more precise knowledge about $p_t^k$.
Also, the variance interval for $p_t^k$ (not shown) will shorten and shift towards zero,
as the Beta distributions in $\MktN$ will be more concentrated
due to the increase of $\nktz$ to $\nktn$.

Alternatively, when there is conflict between prior and observed data (i.e.\ $s^k_t/n_k \not\in [\yktzl, \yktzu]$),
$\PktN$ instead adopts the so-called `banana shape',
arising from the intervals for $\yktn$  being shifted closer to $s^k_t/n_k$
for lower $\nktn$ values than for higher $\nktn$ values, see Figure~\ref{fig:paramsets} (right).
Overall, this results in a wider $\yktn$ interval compared to the no conflict case, 
reflecting the extra uncertainty due to prior-data conflict.  
In other words, the posterior sets make more cautious probability statements about $p_t^k$, as desired in this scenario.

Based on these shapes and \eqref{eq:nyupdate}, it is possible to deduce the following expressions
for the lower and upper bounds of $\yktn$:
\begin{align}
\begin{aligned}
\min_{\PktN} \yktn &=
 \begin{cases}
 \big(\nktzu \yktzl + s^k_t\big) / \big(\nktzu + n_k\big) & \text{if } s^k_t/n_k \ge \yktzl \\
 \big(\nktzl \yktzl + s^k_t\big) / \big(\nktzl + n_k\big) & \text{if } s^k_t/n_k <   \yktzl
 \end{cases}\,,\\
\max_{\PktN} \yktn &=
 \begin{cases}
 \big(\nktzu \yktzu + s^k_t\big) / \big(\nktzu + n_k\big) & \text{if } s^k_t/n_k \le \yktzu \\
 \big(\nktzl \yktzu + s^k_t\big) / \big(\nktzl + n_k\big) & \text{if } s^k_t/n_k >   \yktzu
 \end{cases}\,.
\end{aligned}
\label{eq:ysetupdate}
\end{align}
Note that the lower bound for $\yktn$ is always attained at $\yktzl$, the upper bound at $\yktzu$.
Also note that when $s^k_t/n_k \in [\yktzl, \yktzu]$,
both the lower and the upper bounds for $\yktn$ are attained at $\nktzu$,
corresponding to the spotlight shape.
However, when $s^k_t/n_k \not\in [\yktzl, \yktzu]$,
the banana shape indicates that
one of the bounds for $\yktn$ is attained at $\nktzl$.

\begin{figure}
\includegraphics[width=\textwidth]{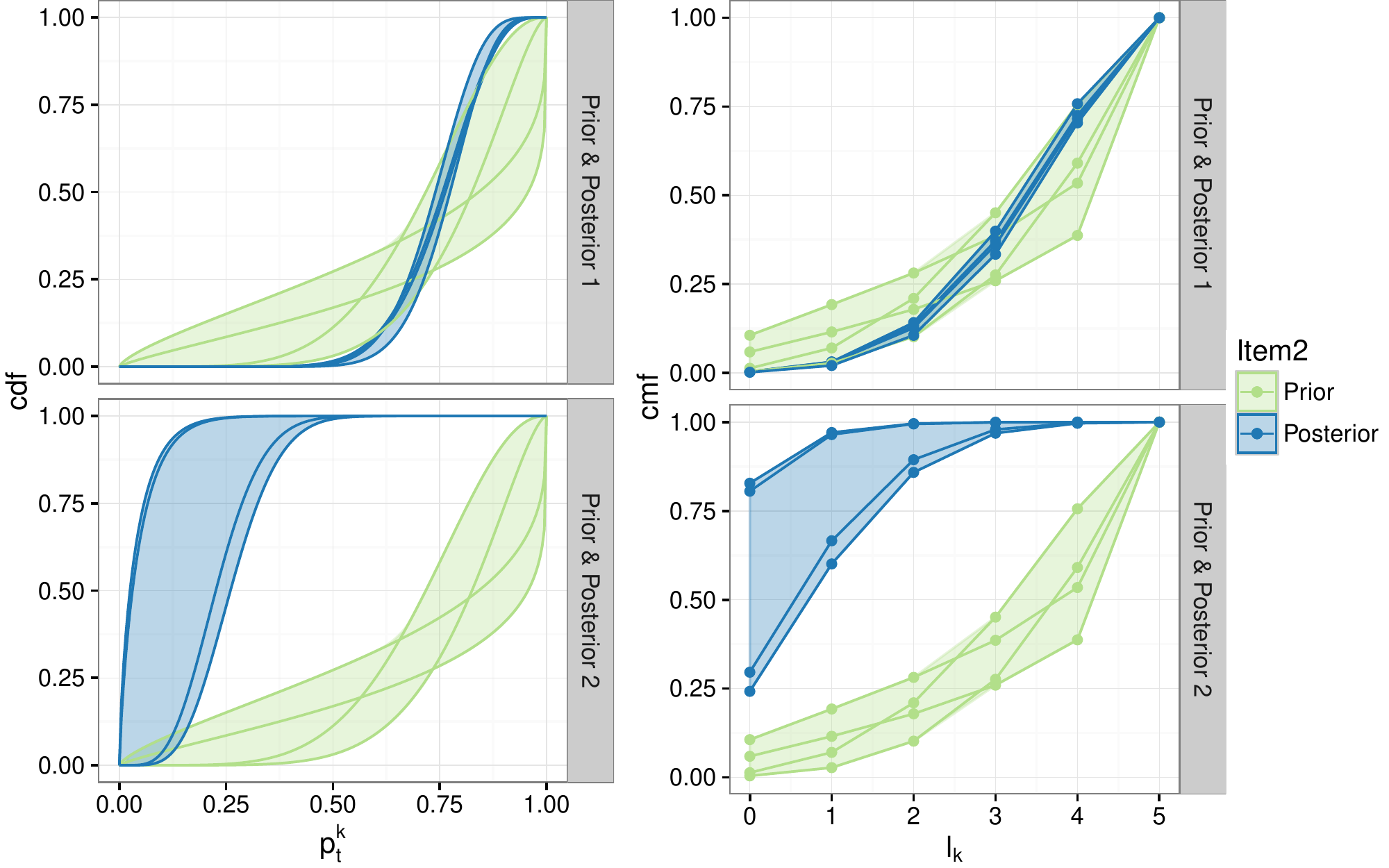}
\caption{Sets of Beta pdfs (left) and Beta-Binomial cmfs (right, for $m_k=5$)
corresponding to the prior and posterior parameter sets in Figure~\ref{fig:paramsets}.
The sets are depicted as shaded areas,
with the distributions corresponding to the four corners
of the prior parameter set $\PktZ$ (or their posterior counterparts) as solid lines.
The top row depicts the set of prior cdfs/cmfs and the set of posterior cdfs/cmfs for the case where data confirm prior assumptions
(see left panel of Figure~\ref{fig:paramsets});
the bottom row depicts the (identical) set of prior cdfs/cmfs and the set of posterior cdfs/cmfs in case of prior-data conflict
(see right panel of Figure~\ref{fig:paramsets}).
The set of posterior cdfs and cmfs
is much larger in case of prior-data conflict:
uncertainty due to this conflict is reflected through increased imprecision.}
\label{fig:betaset-binomset}
\end{figure}

The different locations and sizes of $\PktN$ in the conflict versus no conflict case are then, in turn,
also reflected in the corresponding sets of Beta cdfs and Beta-Binomial cmfs.
As an example, those corresponding to the parameter sets in Figure~\ref{fig:paramsets}
are depicted in Figure~\ref{fig:betaset-binomset}.

In the no conflict case (Posterior 1, top row),
the reduction of the $\yktn$ range in $\PktN$ leads to a much smaller set
of Beta and Beta-Binomial distributions.
For example, the range of predictive probabilities
that two out of a set of five components of type $k$ function at time $t$ 
has changed from $[0.10, 0.28]$ \emph{a priori} to $[0.11,0.14]$ \emph{a posteriori}.
This reflects the gain in precision due to test data in accordance with prior assumptions.

In contrast, for the prior-data conflict case (Posterior 2, bottom row),
the wide $\yktn$ range in $\PktN$ leads to a set of Beta and Beta-Binomial distributions
that is much larger than in the no conflict case.
Here, the range of posterior predictive probabilities
that two out of a set of five components of type $k$ function at time $t$
is now $[0.86, 1.00]$ \emph{a posteriori}, i.e., less precise than in the no conflict case.
Using sets of Beta priors, the resulting set of posterior predictive Beta-Binomial distributions
reflects the precision of prior information,
the amount of data, and prior-data conflict.

Furthermore, with sets of Beta priors it is also possible
to express prior ignorance
by letting $\yktzl \to 0$ and $\yktzu \to 1$
for some or all $t \in {\cal T}$.
(Note that it is not advisable to choose $\yktzl = 0$ and $\yktzu = 1$,
as this can lead to improper posterior predictive distributions.
For example, at any $t < \min(\vec{t}^k)$,
we would have $\yktnu = 1$, leading to one argument of the Beta function
in the denominator of \eqref{eq:postpredCny} being zero.)
These limits for $\yktz$ imply we are only prepared to give trivial bounds for the functioning probability
and do not wish to commit to any specific knowledge about $p^k_t$ \emph{a priori}.
This provides a more natural choice of `noninformative' prior over $[0,1]$ than the usual choice of
a Beta prior with $\alpha^k_t = \beta^k_t = 1$ (or $\nktz = 2$, $\yktz = 0.5$).
Such a prior for all $t \in {\cal T}$ actually reflects a belief that the component reliability function
is on average 1/2 for all $t$, which is not an expression of ignorance, but rather a very specific (and arguably peculiar) prior belief.

%
In a near-noninformative setting, the choice of $\nktzl$ is not relevant,
because \eqref{eq:ysetupdate} implies both lower and upper bound for $\yktn$ are obtained with $\nktzu$.
In particular, $\yktzl > 0$ and $\yktzu < 1$ can be chosen such that
$\frac{s^k_t}{n_k} \in [\yktzl, \yktzu]$ for all $t \in \big(\min(\vec{t}^k), \max(\vec{t}^k)\big)$.
Naturally, one cannot have prior-data conflict in cases of near prior ignorance.

\section{Sets of System Reliability Functions}
\label{sec:setsofrel}

The elements reviewed and extended above culminate hereinafter in the primary contribution of the current work,
providing a framework in which the nonparametric Bayesian system reliability approach developed in \cite{2015:bayessurvsign}
is extended to sets of system reliability functions
by incorporating the sets of priors approach of \citet{2009:WalterAugustin}.
This allows for partial or vague specification of prior component reliability functions,
and enables diagnosis of prior-data conflict which is consequential at the system level.

\subsection{Computation of bounds}

To obtain the lower and upper bound for the system reliability function $\Rsys(t)$,
we now need to minimise and maximise Equation~\eqref{eq:rsyswithsurvsign} over $\PtZi{1}, \ldots, \PtZi{K}$ for each $t$,
where the posterior predictive probabilities for $C^k_t$ are given by the Beta-Binomial pmf \eqref{eq:postpredCny}.
We therefore have
%
\begin{align}
\lefteqn{\lRsys(t \mid \vec{t}^1, \ldots, \vec{t}^K)} \nonumber \\
 &= \min_{\PtZi{1}, \ldots, \PtZi{K}} \Rsys(t \mid \PtZi{1}, \ldots, \PtZi{K}, \vec{t}^1, \ldots, \vec{t}^K) \nonumber \\
 &= \min_{\PtZi{1}, \ldots, \PtZi{K}} 
    \sum_{l_1=0}^{m_1} \cdots \sum_{l_K=0}^{m_K} \Phi(l_1, \ldots, l_K)
                                                 \prod_{k=1}^K P(C^k_t = l_k \mid \yktz, \nktz, s^k_t) \nonumber \\
 &= \min_{\PtZi{1}, \ldots, \PtZi{K}} 
    \sum_{l_1=0}^{m_1} \cdots \sum_{l_K=0}^{m_K} \Phi(l_1, \ldots, l_K) \times \nonumber \\ & \hspace*{12ex}
    \prod_{k=1}^K {m_k \choose l_k} \frac{B(l_k + \nn_{k,t}\yn_{k,t}, m_k - l_k + \nn_{k,t}(1-\yn_{k,t}))}
                                         {B(\nn_{k,t}\yn_{k,t}, \nn_{k,t}(1-\yn_{k,t}))} \nonumber \\
 &= \min_{\PtZi{1}, \ldots, \PtZi{K}} 
    \sum_{l_1=0}^{m_1} \cdots \sum_{l_K=0}^{m_K} \Phi(l_1, \ldots, l_K) \times \nonumber \\ & \hspace*{8ex}
    \prod_{k=1}^K {m_k \choose l_k} \frac{B(l_k + \nktz\yktz + s^k_t, m_k - l_k + \nktz(1-\yktz) + n_k - s^k_t)}{B(\nktz\yktz + s^k_t, \nktz(1-\yktz) + n_k - s^k_t)}
    \,, \label{eq:LwrSysPostA}
\end{align}
and similarly maximising for $\uRsys(\cdot)$.

Note that $\Phi(\cdot)$ is non-decreasing in each of its arguments $l_1,\ldots,l_K$,
thus if there is first-order stochastic ordering on $P(C^k_t = l_k \mid \yktz, \nktz, s^k_t)$
for each $k$, then this ordering can be used to determine the elements of $\PtZi{k}$
which minimise and maximise the overall system reliability function without
resorting to computationally expensive exhaustive searches or numerical optimisation.

We therefore start by providing the following result, where indices are suppressed for readability.
We use $\ge_{\mathrm{st}}$ to denote first-order stochastic dominance.

\begin{theorem}
  \label{thm:y}
  Let $\beta_y$ denote the Beta-Binomial distribution with probability mass function parameterised as:
  \[ p(l \mid y, n, m, s, N) \propto \frac{B(l + ny + s, m - l + n(1-y) + N - s)}{B(ny + s, n(1-y) + N - s)}, \]
  with $n, m, s,$ and $N$ fixed and unknown.
  
  Then $\beta_{\yu} \ge_{\mathrm{st}} \beta_{\yl} \ \forall \ \yu > \yl$ with $\yu, \yl \in (0,1)$.
\end{theorem}
The proof is provided in Appendix \ref{ap:proofs}, p.\pageref{prf:y}.

Consequently, for each component, the posterior predictive Beta Binomial distributions 
with larger prior functioning probability stochastically dominate those
with smaller prior functioning probability, providing rigorous proof 
which accords with intuition. Applying this result to the sets of system reliability 
functions, together with the monotonicity in the survival signature, means that
$\lRsys(\cdot)$ is attained when $\yktz = \yktzl$ and $\uRsys(\cdot)$
is attained when $\yktz = \yktzu$ for all possible $\nktz$ values.

The analogous result for $\nktz$ is more subtle, because stochastic 
dominance is not guaranteed at a single value.  The following Theorem 
provides simple sufficient conditions under which an upper or lower 
limit has first-order stochastic dominance and has virtually no 
computational overhead to test.

\begin{theorem}
  \label{thm:n}
  Let $\beta_n$ denote the Beta-Binomial distribution with probability mass function parameterised as:
  \[ p(l \mid y, n, m, s, N) \propto \frac{B(l + ny + s, m - l + n(1-y) + N - s)}{B(ny + s, n(1-y) + N - s)}, \]
  with $y, m, s,$ and $N$ fixed and unknown.  
  Then,
  \[ y > \frac{s + m - 1}{N + m - 1} \implies \beta_{\nu} \ge_{\mathrm{st}} \beta_{\nl} \]
  and 
  \[ y < \frac{s}{N + m - 1} \implies \beta_{\nu} \le_{\mathrm{st}} \beta_{\nl} \]
\end{theorem}
The proof is provided in Appendix \ref{ap:proofs}, p.\pageref{prf:n}.

If $\frac{s}{N + m - 1} < y < \frac{s + m - 1}{N + m - 1}$ then 
Theorem 2 cannot determine stochastic dominance.  The following Lemma which is slightly
more computationally costly, but still much faster than an exhaustive search, may be able to determine first-order
stochastic dominance in such situations.
\begin{lemma}
  \label{lem:n}
  Let $\beta_n$ denote the Beta-Binomial distribution as in Theorem \ref{thm:n}.  Define
  \[ \mathcal{L}_{\nu,\nl}(l) := \frac{p(l \mid y, \nu, m, s, N)}{p(l \mid y, \nl, m, s, N)} \]
  Then,
  \[ \left. \begin{aligned}
       \mathcal{L}_{\nu,\nl}(0) \le 1 \\
       \mathcal{L}_{\nu,\nl}(m) \ge 1 \\
    \end{aligned} \right\}
    \implies \beta_{\nu} \ge_{\mathrm{st}} \beta_{\nl} \]
  and
  \[ \left. \begin{aligned}
       \mathcal{L}_{\nu,\nl}(0) \ge 1 \\
       \mathcal{L}_{\nu,\nl}(m) \le 1 \\
    \end{aligned} \right\}
    \implies \beta_{\nu} \le_{\mathrm{st}} \beta_{\nl} \]
\end{lemma}
The proof is provided in Appendix \ref{ap:proofs}, p.\pageref{prf:n2}.
In the cases where neither Theorem \ref{thm:n} or Lemma \ref{lem:n} apply,
the entire posterior system reliability function must be optimised to 
find the minima/maxima.
In practice, in the examples to be presented in the sequel, Theorem 
\ref{thm:n} and Lemma \ref{lem:n} do provide guarantees of first-order
stochastic dominance for the vast majority of time points, $t$,
substantially lowering the computational costs of performing the
minimisation/maximisation involved in finding the sets of system 
reliability functions compared to either numerical optimisation or
an exhaustive grid search (which would get exponentially slower in 
the number of different components).

Thus,
\begin{align}
\lefteqn{\lRsys(t \mid \vec{t}^1, \ldots, \vec{t}^K)} \nonumber \\
 &= \min_{\PtZi{1}, \ldots, \PtZi{K}} 
    \sum_{l_1=0}^{m_1} \cdots \sum_{l_K=0}^{m_K} \Phi(l_1, \ldots, l_K) \times \nonumber \\ & \hspace*{8ex}
    \prod_{k=1}^K {m_k \choose l_k} \frac{B(l_k + \nktz\yktz + s^k_t, m_k - l_k + \nktz(1-\yktz) + n_k - s^k_t)}{B(\nktz\yktz + s^k_t, \nktz(1-\yktz) + n_k - s^k_t)}
    \nonumber\\
 &= \sum_{l_1=0}^{m_1} \cdots \sum_{l_K=0}^{m_K} \Phi(l_1, \ldots, l_K) \times \nonumber \\ & \hspace*{8ex}
    \prod_{k=1}^K {m_k \choose l_k} \frac{B(l_k + \nktzo\yktzl + s^k_t, m_k - l_k + \nktzo(1-\yktzl) + n_k - s^k_t)}{B(\nktzo\yktzl + s^k_t, \nktzo(1-\yktzl) + n_k - s^k_t)} \label{eq:LwrSysPost}\\
 &\mathrm{where} \nonumber \\
 &\nktzo = \left\{ \begin{aligned}
   \nktzu & \quad\mbox{if}\ \yktzl < \frac{s^k_t}{n_k + m_k - 1} \vee \Big( \mathcal{L}_{\nktzu,\nktzl}(0) \ge 1 \wedge \mathcal{L}_{\nktzu,\nktzl}(m) \le 1 \Big) \\
   \nktzl & \quad\mbox{if}\ \yktzl > \frac{s^k_t + m_k - 1}{n_k + m_k - 1} \vee \Big( \mathcal{L}_{\nktzu,\nktzl}(0) \le 1 \wedge \mathcal{L}_{\nktzu,\nktzl}(m) \ge 1 \Big) \\
   & \quad\mbox{optimised otherwise}
 \end{aligned} \right. \nonumber
\end{align}

The result for $\uRsys(\cdot)$ is completely analogous.  It is interesting
to note that if $m_k=1$ the bounds are sharp on stochastic dominance.
In particular, when $m_k=1$, $\yktzl < \frac{s^k_t}{n_k}$ indicates the 
lower bound is not in conflict with the observed data, whilst
$\yktzl > \frac{s^k_t}{n_k}$ is in conflict since the observed
empirical probability of functioning at time $t$ is below the
prior lower bound.  Consequently, note that $\nktzl$ is 
used only when the prior comes into conflict with the data.
Since $\nktz$ controls the prior certainty, this accords with
the intuition that the least certain prior bound is invoked
when in a conflict setting and the more certain prior bound used when the data agrees.

\subsection{Prior parameter choice}

In the following, we will give some guidelines on how to choose the parameter sets $\PkZi{1}, \ldots, \PkZi{\tmax}$
which define the set of prior discrete reliability functions for components of type $k$.
We advocate that this is much easier in terms of $\nz$ and $\yz$ than it would be in terms of $\alpha$ and $\beta$.

As mentioned in Section~\ref{sec:nonparamapproach}, the functioning probabilities $\ptk$
must satisfy $p^k_{t_j} \ge p^k_{t_{j+1}}$.
This naturally translates to conditions on the prior for $\ptk$,
so that for example 
$\yzu_{k,t_j} \ge \yzu_{k,t_{j+1}}$ and $\yzl_{k,t_j} \ge \yzl_{k,t_{j+1}}$ should hold.
Because $s^k_t/n_k$ is decreasing in $t$, the weighted average property of the update step in
Equation \eqref{eq:nyupdate} for $\ykt$ ensures that 
$\ynu_{k,t_j} \ge \ynu_{k,t_{j+1}}$ and $\ynl_{k,t_j} \ge \ynl_{k,t_{j+1}}$.
In situations where one has a high degree of certainty about the functioning probability for low $t$,
but less certainty about what happens for larger $t$,
then one can let $\yktzl$ drop to (almost) 0, but clearly $\yktzu$ should not increase.

It is inadvisable to express certainty in the expected functioning probabilities
with $\nktz$ bounds that vary substantially over the range of $t$.
With (strongly) differing $\nktz$ bounds, monotonicity of the $\yktn$ bounds cannot be guaranteed.
For example, if $\yzu_{k,t_j} = \yzu_{k,t_{j+1}}$, $\yzl_{k,t_j} = \yzl_{k,t_{j+1}}$,
and $s^k_{t_j}/n_k \in [\yzl_{k,t_j}, \yzu_{k,t_j}]$ (meaning there is no prior-data conflict),
then should there be no observed failures in $[t_j, t_{j+1}]$, so that $s^k_{t_{j+1}}/n_k = s^k_{t_{j}}/n_k$, then
\begin{align*}
  \nzu_{k,t_j} < \nzu_{k,t_{j+1}} \quad&\implies\quad\ynu_{k,t_j} < \ynu_{k,t_{j+1}} \quad\mbox{and} \\
  \nzu_{k,t_j} > \nzu_{k,t_{j+1}} \quad&\implies\quad\ynl_{k,t_j} < \ynl_{k,t_{j+1}}
\end{align*}
Again, this follows from \eqref{eq:nyupdate}, the weighted average property.
It is possible to construct similar examples with regard to the lower bound $\nzl_{k,t_j}$.
Therefore, we advise taking the same $\nktz$ bounds for all $t$ as far as possible.
If they do change, it must be very gradual and we recommend diagnosing 
any problems as above.

Generally, the interpretation as pseudocount or prior strength should guide the choice of bounds for $\nktz$; 
low values for $\nktz$ as compared to the test sample size $n_k$
give low weight to the prior expected functioning probability intervals $[\yktzl, \yktzu]$,
and the location of posterior intervals $[\yktnl, \yktnu]$ will be dominated by the location of $s^k_t/n_k$.
Furthermore, the length of $[\yktnl, \yktnu]$ is shorter for low $\nktz$ values.  Specifically, 
in a no-conflict situation, when $\nktzu = n_k$ then $[\yktnl, \yktnu]$ has half the length of $[\yktzl, \yktzu]$.
In contrast, high values for $\nktz$ will lead to slower learning and wider $\yktn$ intervals,
which means more cautious posterior inferences.
The difference betweeen $\nktzu$ and $\nktzl$ determines the strength of the prior-data conflict sensitivity;
as is clear from Figure~\ref{fig:paramsets} and \eqref{eq:ysetupdate}, the wider the $\nktz$ interval,
the wider $[\yktnl, \yktnu]$ in case of conflict.
So it seems useful to choose $\nktzl = 1$ or $\nktzl = 2$,
while choosing $\nktzu$ with help of the half-width rule as described above.

As mentioned in Section~\ref{sec:setsofbetapriors},
it is not advisable to choose $\yktzl = 0$ and $\yktzu = 1$.
For any $t \not\in \big(\min(\vec{t}^k), \max(\vec{t}^k)\big)$,
this can lead to improper posterior predictive distributions.
However, it is possible to choose values close to $0$ and $1$, respectively,
and due to the linear update step \eqref{eq:nyupdate} for $\yktn$,
posterior inferences are not overly sensitive to whether $\yktnu = 0.99$ or $\yktnu = 0.9999$.
Likewise, our nonparametric method does not cause unintuitive tail behaviour
as some parametric methods do;
there is no problem, for example, with assigning $\yktnu$ near-zero for large $t$ if prior knowledge suggests so.

While it is possible to set the bounds $\yktzl$ and $\yktzu$ for each $t \in {\cal T}$ individually,
in practice this will be often too time-consuming when ${\cal T}$ forms a dense grid.
Switching to a coarser time grid will waste information from data,
as then failure times in the test data are rounded up to the next $t \in {\cal T}$.
In the examples here we elicit bounds for a subset of ${\cal T}$
and fill up the time grid with the least committal bounds, i.e.,
taking $\yktzu$ equal to last (in the time sequence) elicited $\yktzu$,
and likewise $\yktzl$ equal to next (in time sequence) elicited $\yktzl$.
A possible elicitation procedure in this vein could be
to start with eliciting $\yktz$ bounds for a few `central' time points $t$,
filling up the grid as described above accordingly,
and then to further refine the obtained bounds as deemed necessary by the expert.

\section{Practical Usage and Examples}
\label{sec:examples}

\subsection{Software}

The methods of this paper have been implemented in the \textbf{R}
\citep{R} package \texttt{ReliabilityTheory} \citep{2015:aslett-RT},
providing an easy to use interface for reliability practitioners.  The
primary function, which computes the upper and lower posterior
predictive system survival probabilities as in \eqref{eq:LwrSysPost}, is
named \texttt{nonParBayesSystemInferencePriorSets()}.  The user
specifies the times at which to evaluate the bounds, the survival
signature ($\Phi(\cdot)$), the component test data ($\vec{t}^1, \ldots,
\vec{t}^K$), and the prior parameter set for each component type and
time ($\PtZi{k}$, via $\nzu_{k,t}, \nzl_{k,t}, \yzu_{k,t}$, and
$\yzl_{k,t}$).  All computations of $\lRsys$
and $\uRsys$ at different time points are performed in parallel
automatically where the CPU has multiple cores and making automatic
use of the theoretical results in Section \ref{sec:setsofrel} where
applicable, performing exhaustive search in the few cases they are not.

Note that computation of the system signature itself can be simplified by
expressing the structure of the system as an undirected graph using the
\texttt{computeSystemSurvivalSignature()} function in the same package,
leaving only data and prior to be handled.  These publicly available
functions have been used in computing all the following examples for
reproducibility.  See Appendix \ref{ap:software} for further details of how to use this software.

\subsection{Examples}

\subsubsection{Toy example}

\begin{figure}
\centering
\begin{tikzpicture}
[type1/.style={rectangle,draw,fill=black!20,very thick,inner sep=0pt,minimum size=8mm},
 type2/.style={rectangle,draw,fill=black!20,very thick,inner sep=0pt,minimum size=8mm},
 type3/.style={rectangle,draw,fill=black!20,very thick,inner sep=0pt,minimum size=8mm},
 cross/.style={cross out,draw=red,very thick,minimum width=7mm, minimum height=5mm},
 hv path/.style={thick, to path={-| (\tikztotarget)}},
 vh path/.style={thick, to path={|- (\tikztotarget)}}]
\node[type1] (T3)   at ( 5.6, 0) {T3};
\node[type2] (T2)   at ( 2.8, 0) {T2};
\node[type3] (T1-2) at ( 1.4, 1) {T1};
\node[type3] (T1-3) at ( 1.4,-1) {T1};
\node[type3] (T1-5) at ( 4.2, 1) {T1};
\node[type3] (T1-6) at ( 4.2,-1) {T1};
\coordinate (start) at (0  ,0);
\coordinate (end)   at (6.3,0);
\coordinate (bista) at (0.7,0);
\coordinate (biend) at (4.9,0);
\path (bista)     edge[hv path] (start)
                  edge[vh path] (T1-2.west)
                  edge[vh path] (T1-3.west)
      (T1-2.east) edge[hv path] (T2.north)
      (T1-3.east) edge[hv path] (T2.south)
      (T2.north)  edge[vh path] (T1-5.west)
      (T2.south)  edge[vh path] (T1-6.west)
      (biend)     edge[vh path] (T1-5.east)
                  edge[vh path] (T1-6.east)
                  edge[hv path] (T3.west)
      (T3.east)   edge[hv path] (end);
\end{tikzpicture}
\caption{Reliability block diagram for a `bridge' system with three component types.}
\label{fig:bridge-layout}
\end{figure}
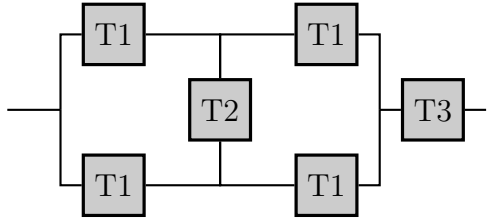

\begin{figure}
\includegraphics[width=\textwidth]{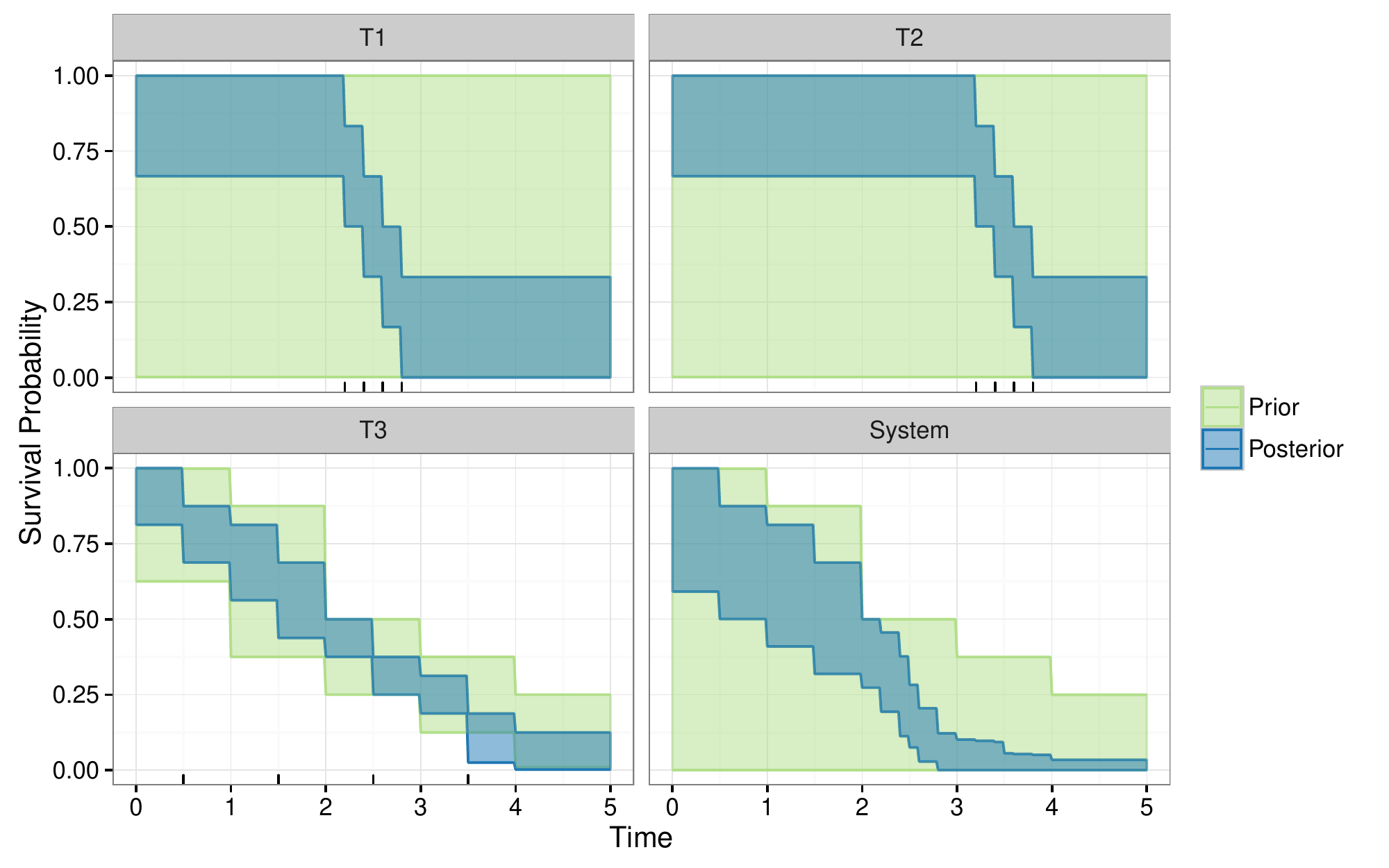}
\caption{Prior and posterior sets of reliability functions for the `bridge' system and its three component types,
with failure times as expected for component type T3.
Test data failure times are denoted with tick marks near the time axis.}
\label{fig:bridge-fitting}
\end{figure}

\begin{figure}
\includegraphics[width=\textwidth]{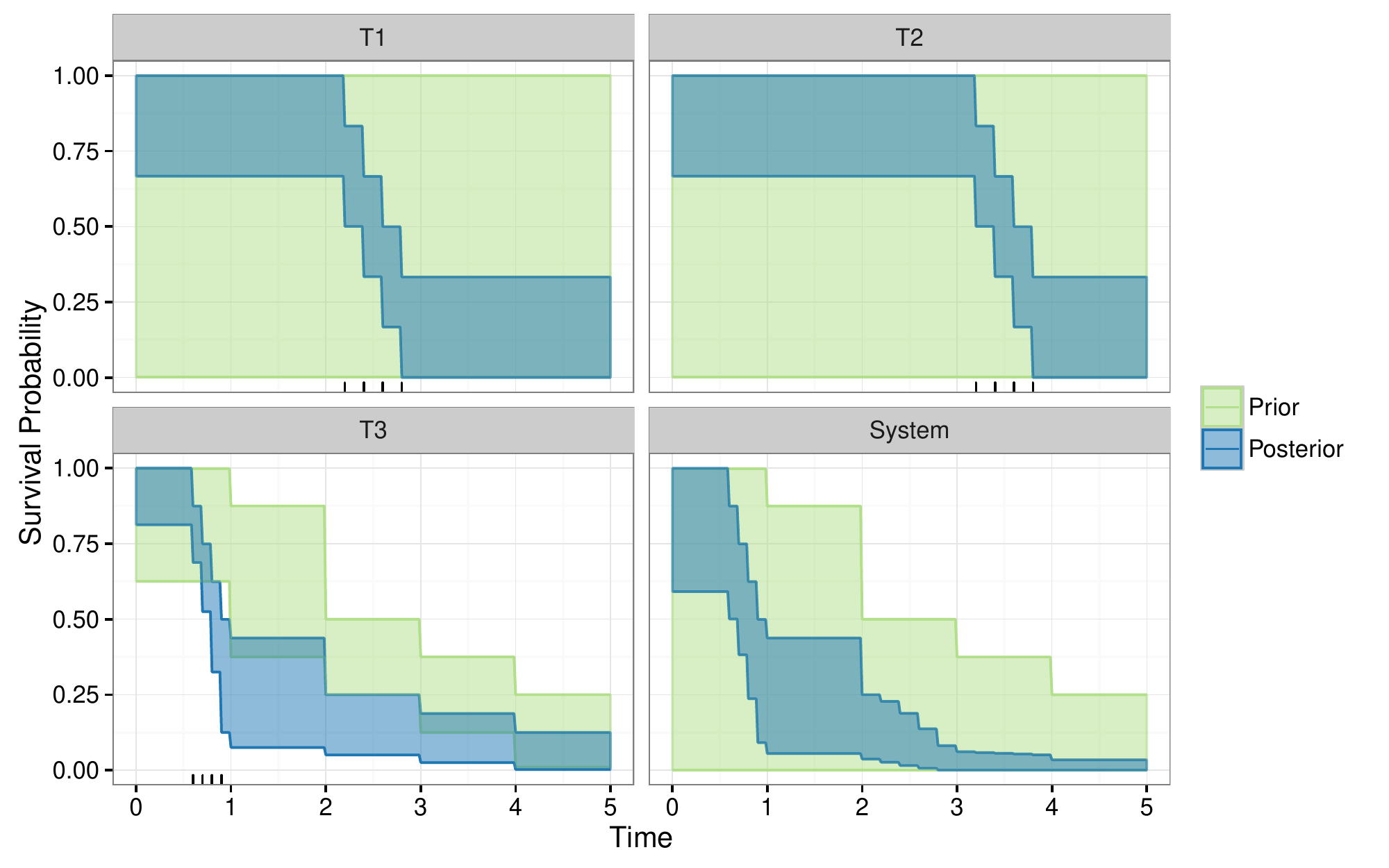}
\caption{Prior and posterior sets of reliability functions for the `bridge' type system and its three component types,
with failure times earlier as expected for component type T3.}
\label{fig:bridge-early}
\end{figure}

\begin{figure}
\includegraphics[width=\textwidth]{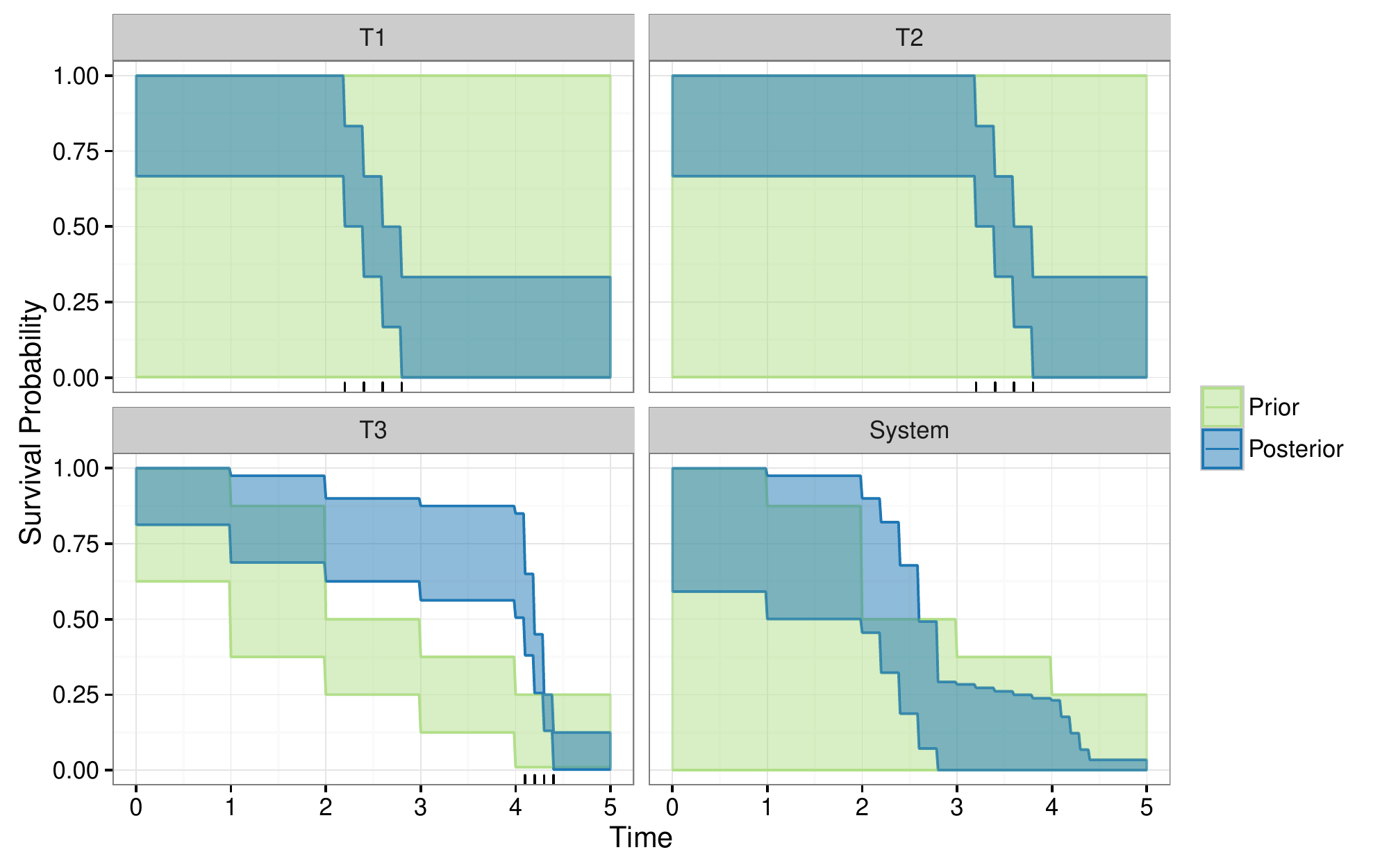}
\caption{Prior and posterior sets of reliability functions for the `bridge' type system and its three component types,
with failure times later as expected for component type T3.}
\label{fig:bridge-late}
\end{figure}

As a toy example, consider a `bridge' type system layout with three types of components T1, T2 and T3,
as depicted in Figure~\ref{fig:bridge-layout}.
The survival signature for this system is given in Table~\ref{tab:bridge-survsign}.
All rows with T3 $=0$ have been omitted; without T3, the system cannot function, thus $\Phi = 0$.
\begin{table}
\centering
\begin{tabular}{ccclccccl}
  \toprule
T1 & T2 & T3 & $\Phi$ & \quad & T1 & T2 & T3 & $\Phi$\\ 
  \midrule
0 & 0 & 1 & 0    & & 0 & 1 & 1 & 0 \\ 
1 & 0 & 1 & 0    & & 1 & 1 & 1 & 0 \\ 
2 & 0 & 1 & 0.33 & & 2 & 1 & 1 & 0.67 \\ 
3 & 0 & 1 & 1    & & 3 & 1 & 1 & 1 \\ 
4 & 0 & 1 & 1    & & 4 & 1 & 1 & 1 \\ 
   \bottomrule
\end{tabular}
\caption{Survival signature for the bridge system from Figure~\ref{fig:bridge-layout},
omitting all rows with T3 $=0$, since $\Phi=0$ for these.}
\label{tab:bridge-survsign}
\end{table}
For component types T1 and T2, we consider a near-noninformative set of prior reliability functions.
For components of type T3, we consider an informative set of prior reliability functions
as given in Table~\ref{tab:bridge-T3prior}.
This set could result from eliciting prior functioning probabilities at times $0,1,2,3,4,5$ only,
and filling up the rest.
These prior assumptions, together with sets of posterior reliability functions
resulting from three different scenarios for test data for component type T3,
are illustrated in Figures~\ref{fig:bridge-fitting}, \ref{fig:bridge-early} and \ref{fig:bridge-late}; 
test data for components of type T1 and T2 are invariably taken as
$\vec{t}^1 = (2.2, 2.4, 2.6, 2.8)$ and $\vec{t}^2 = (3.2, 3.4, 3.6, 3.8)$, respectively.

\begin{table}
\centering
\begin{tabular}{crrrrr}
\toprule
$t$ & $[0,1)$ & $[1,2)$ & $[2,3)$ & $[3,4)$ & $[4,5)$ \\
\midrule
$\ytzl{3}$ & 0.625 & 0.375 & 0.250 & 0.125 & 0.010 \\
$\ytzu{3}$ & 0.999 & 0.875 & 0.500 & 0.375 & 0.250 \\
\bottomrule
\end{tabular}
\caption{Lower and upper prior functioning probability bounds for component type T3 in the `bridge' system example.}
\label{tab:bridge-T3prior}
\end{table}

In Figure~\ref{fig:bridge-fitting}, 
test data for component type T3 is $\vec{t}^3 = (0.5, 1.5, 2.5, 3.5)$,
and so in line with expectations.
The posterior set of reliability functions for each component type and the whole system
is considerably smaller compared to the prior set
(due to the low prior strength intervals
$[\ntzl{1},\ntzu{1}] = [\ntzl{2},\ntzu{2}] = [1,2]$, $[\ntzl{3},\ntzu{3}] = [1,4]$)
and so giving more precise reliability statements.
We see that posterior lower and upper functioning probabilities drop at those times $t$
when there is a failure time in the test data,
or a drop in the prior functioning probability bounds.
Note that the lower bound for the prior system reliability function is zero
due to the prior lower bound of zero for T1;
for the system to function, at least two components of type T1 must function.

In Figure~\ref{fig:bridge-early},
test data of component type T3 is $\vec{t}^3 = (0.6, 0.7, 0.8, 0.9)$,
and so earlier than expected.
Compared to Figure~\ref{fig:bridge-fitting},
posterior functioning intervals for T3 are wider between $t=1$ and $t=3.5$,
reflecting additional imprecision due to prior-data conflict.
For $t > 1$, it is clearly visible how $\ytnu{3}$ is halfway between $\ytzu{3}$ and $s_t^3/n_3 = 0$
(weights $\ntzu{3} = 4$ and $n_3 = 4$),
while $\ytnl{3}$ is one-fifth of $\ytzl{3}$
(weights $\ntzl{3} = 1$ and $n_3 = 4$).
Note that the posterior system functioning probability is constant for $t \in [1,2]$
because in that interval the prior functioning probability is constant
and there are no observed failures.

In Figure~\ref{fig:bridge-late}, 
test data of component type T3 is $\vec{t}^3 = (4.1, 4.2, 4.3, 4.4)$,
and so observed failures are later than expected.
Here we see that for $t \in [2,4]$,
posterior functioning bounds for T3 are even wider than prior functioning bounds.
The width turns back to being half the prior width
only after the four failures.
The imprecision carries over to the system bounds,
where we see wider bounds as compared to the other two scenarios
especially between $t=2$ and $t=4$. In particular, also note
that at the system level posterior bounds are a subset of prior bounds after
$t = 2.6$, although prior-data conflict for the 
individual component type T3 extends well beyond $t=4$.  This 
demonstrates the power of this technique to identify prior-data conflict
which is actually consequential at the system 
level, not just the component level --- in other words, for 
mission times $t>2.6$, we can diagnose that the prior-data conflict 
need not be of elevated concern for this system viewed as a whole.
Nevertheless, the posterior system reliability bounds are wider than in the no-conflict case for $t \in [1, 4.4]$,
signalising the general need for caution in this scenario.

\subsubsection{Automotive brake system}

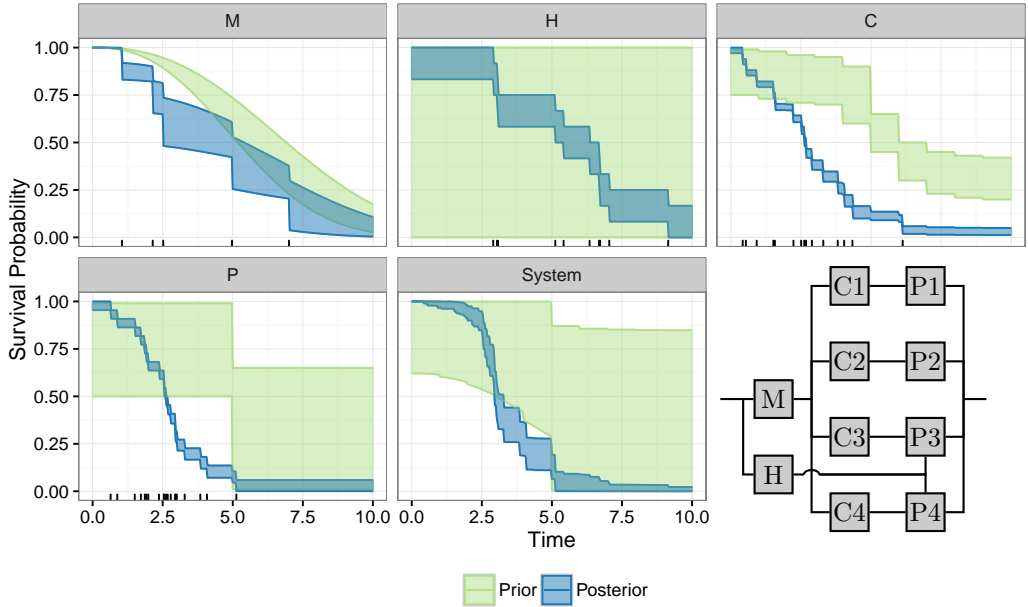
\begin{figure}
\begin{tikzpicture}
[typeM/.style={rectangle,draw,fill=black!20,thick,inner sep=0pt,minimum size=5mm,font=\footnotesize},
 typeC/.style={rectangle,draw,fill=black!20,thick,inner sep=0pt,minimum size=5mm,font=\footnotesize},
 typeP/.style={rectangle,draw,fill=black!20,thick,inner sep=0pt,minimum size=5mm,font=\footnotesize},
 typeH/.style={rectangle,draw,fill=black!20,thick,inner sep=0pt,minimum size=5mm,font=\footnotesize},
 cross/.style={cross out,draw=red,very thick,minimum width=7mm, minimum height=5mm},
 hv path/.style={thick, to path={-| (\tikztotarget)}},
 vh path/.style={thick, to path={|- (\tikztotarget)}}]
\node at (0,0) {\includegraphics[width=\textwidth]{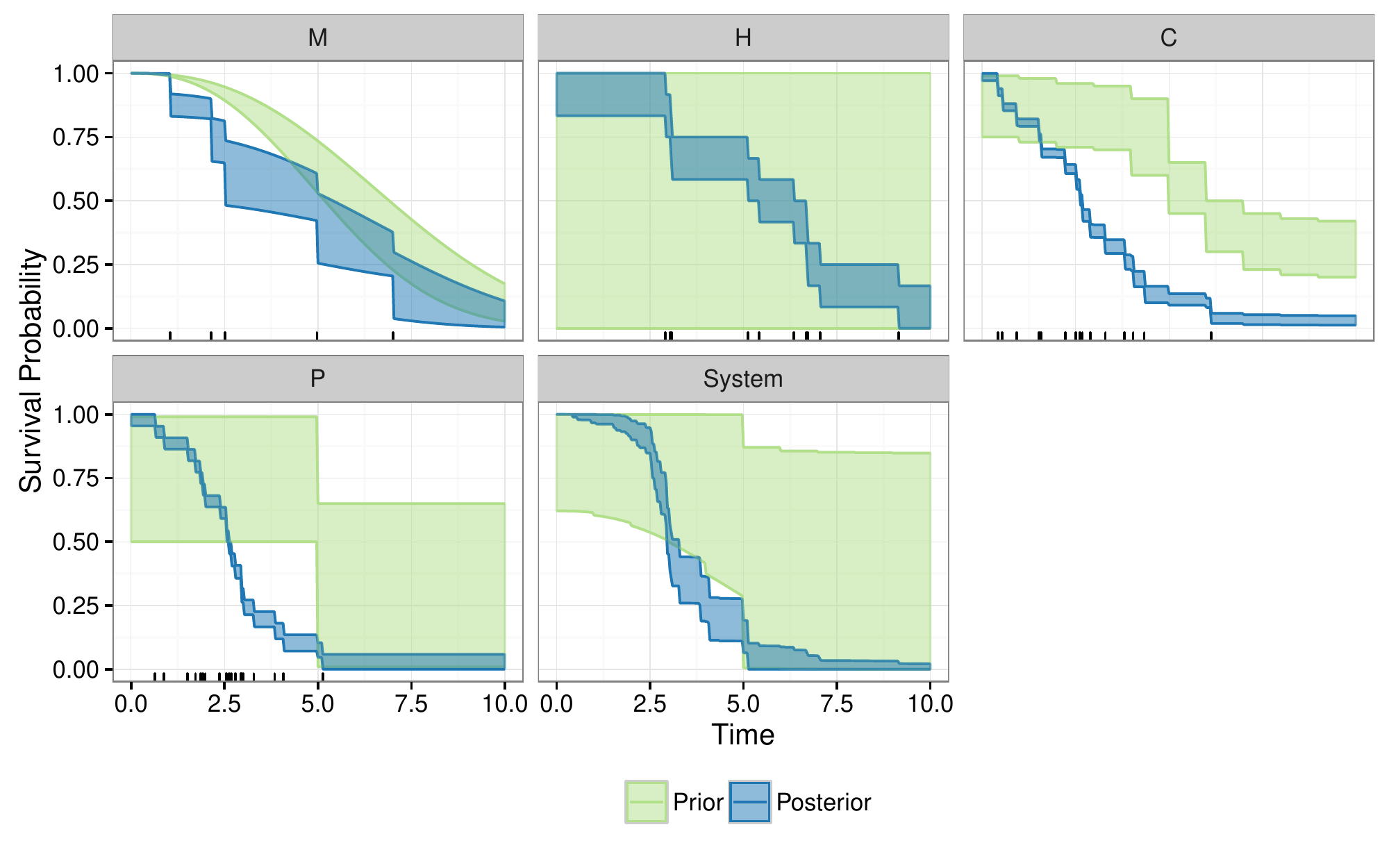}};
\begin{scope}[scale=1.00,xshift=3.4cm,yshift=-1.1cm]
\node[typeM] (M)    at ( 0  , 0  ) {M};
\node[typeC] (C1)   at ( 1  , 1.5) {C1};
\node[typeC] (C2)   at ( 1  , 0.5) {C2};
\node[typeC] (C3)   at ( 1  ,-0.5) {C3};
\node[typeC] (C4)   at ( 1  ,-1.5) {C4};
\node[typeP] (P1)   at ( 2  , 1.5) {P1};
\node[typeP] (P2)   at ( 2  , 0.5) {P2};
\node[typeP] (P3)   at ( 2  ,-0.5) {P3};
\node[typeP] (P4)   at ( 2  ,-1.5) {P4};
\node[typeH] (H)    at ( 0  ,-1  ) {H};
\coordinate (start)  at (-0.7, 0);
\coordinate (startC) at ( 0.5, 0);
\coordinate (startH) at (-0.4, 0);
\coordinate (Hhop1)  at ( 0.4,-1);
\coordinate (Hhop2)  at ( 0.6,-1);
\coordinate (endP)   at ( 2.5, 0);
\coordinate (end)    at ( 2.8, 0);
\path (start)     edge[hv path] (M.west)
      (M.east)    edge[hv path] (startC)
      (startC)    edge[vh path] (C1.west)
                  edge[vh path] (C2.west)
                  edge[vh path] (C3.west)
                  edge[vh path] (C4.west)
      (C1.east)   edge[hv path] (P1.west)
      (C2.east)   edge[hv path] (P2.west)
      (C3.east)   edge[hv path] (P3.west)
      (C4.east)   edge[hv path] (P4.west)
      (endP)      edge[vh path] (P1.east)
                  edge[vh path] (P2.east)
                  edge[vh path] (P3.east)
                  edge[vh path] (P4.east)
                  edge[hv path] (end)
      (startH)    edge[vh path] (H.west)
      (H.east)    edge[hv path] (Hhop1)
      (Hhop1)     edge[thick,out=90,in=90] (Hhop2)
      (Hhop2)     edge[hv path] (P3.south)
                  edge[hv path] (P4.north);
\end{scope}
\end{tikzpicture}
\caption{Prior and posterior sets of reliability functions for a simplified automotive brake system
with layout as depicted in the lower right panel.}
\label{fig:brakesystem}
\end{figure}

We also consider a simplified automotive brake system.
The master brake cylinder (M) activates all four wheel brake cylinders (C1 -- C4),
which in turn actuate a braking pad assembly each (P1 -- P4).
The hand brake mechanism (H) goes directly to the brake pad assemblies P3 and P4;
the car brakes when at least one brake pad assembly is actuated.
All values for $\Phi \not\in \{0,1\}$ are given in Table~\ref{tab:brake-survsign}.
\begin{table}
\centering
\begin{tabular}{cccclcccccl}
  \toprule
M & H & C & P & $\Phi$ & \quad & M & H & C & P & $\Phi$\\ 
  \midrule
1 & 0 & 1 & 1 & 0.25 & & 1 & 0 & 2 & 1 & 0.50 \\ 
1 & 0 & 1 & 2 & 0.50 & & 1 & 0 & 2 & 2 & 0.83 \\ 
1 & 0 & 1 & 3 & 0.75 & & 1 & 0 & 3 & 1 & 0.75 \\ 
0 & 1 & 0 & 1 & 0.50 & & 1 & 1 & 0 & 1 & 0.50 \\ 
0 & 1 & 0 & 2 & 0.83 & & 1 & 1 & 0 & 2 & 0.83 \\ 
0 & 1 & 1 & 1 & 0.62 & & 1 & 1 & 1 & 1 & 0.62 \\ 
0 & 1 & 1 & 2 & 0.92 & & 1 & 1 & 1 & 2 & 0.92 \\ 
0 & 1 & 2 & 1 & 0.75 & & 1 & 1 & 2 & 1 & 0.75 \\ 
0 & 1 & 2 & 2 & 0.97 & & 1 & 1 & 2 & 2 & 0.97 \\ 
0 & 1 & 3 & 1 & 0.88 & & 1 & 1 & 3 & 1 & 0.88 \\ 
   \bottomrule
\end{tabular}
\caption{Survival signature values $\not\in \{0,1\}$ for the simplified automotive brake system depicted in Figure~\ref{fig:brakesystem}.
}
\label{tab:brake-survsign}
\end{table}
The system layout is depicted in Figure~\ref{fig:brakesystem},
together with prior and posterior sets of reliability functions for the four component types and the complete system.
Observed lifetimes from test data are indicated by tick marks in each of the four component type panels,
where $n_\text{M}=5$, $n_\text{H}=10$, $n_\text{C}=15$, and $n_\text{P}=20$.
We assume $[\ntzl{\text{M}},\ntzu{\text{M}}] = [1,8]\ \forall t$,
and $[\nktzl, \nktzu] = [1,2]$ for $k \in \{\text{H, C, P}\}$ and all $t$.
Prior functioning probability bounds for M are based on
a Weibull cdf with shape $2.5$ and scales $6$ and $8$ for the lower and upper bound, respectively.
The prior bounds for P can be seen as the least committal bounds
derived from an expert statement of $\ytz{\text{P}} \in [0.5, 0.65]$ for $t=5$ only.
For H, near-noninformative prior functioning probability bounds have been selected;
with the upper bound for P being approximately one for $t \le 5$ as well,
the prior upper system reliability bound for $t \le 5$ is close to one, too,
since the system can function on H and one of P1 -- P4 alone.
Note that the posterior functioning probability interval for M
is wide not only due to the limited number of observations,
but also because $\ntzu{\text{M}} = 8$ and the prior-data conflict reaction.

Posterior functioning probability bounds for the complete system
are much more precise than the prior system bounds,
reflecting the information gained from component test data.
The posterior system bounds can be also seen to reflect location and precision of the component bounds;
for example, the system bounds drop drastically between $t=2.5$ and $t=3.5$
mainly due to the drop of the bounds for P at that time.

It is also interesting to note that the prior-data conflict 
which is consequential at the system level occurs over 
roughly the same range in $t$ as there is prior-data 
conflict for component type P.
Indeed, this occurs despite there being 
prior-data conflict in both M and C over much larger ranges, 
giving valuable insight into which prior requires further 
expert attention --- thus the technique avoids wasted time 
addressing prior-data conflict in components which may not 
be relevant when propagated to the uncertainty in the whole system.



%

\section{Conclusions}

In this paper we have contributed an imprecise Bayesian nonparametric approach
to system reliability with multiple types of components.
The approach allows modelling partial or imperfect prior knowledge on component failure distributions
in a flexible way through bounds on the functioning probability for a grid of time points, 
and combines this information with test data in an imprecise Bayesian framework.
Component-wise predictions on the number of functioning components
are then combined to bounds for the system survival probability by means of the survival signature.
New results on first-order stochastic dominance for the Beta-Binomial distribution
enable closed-form solutions for these bounds in most cases and 
avoid exponential growth in the complexity of computing the estimate
as the number of components grows.
The widths of the resulting system reliability bounds
reflect the amount of test data, the precision of prior knowledge,
and crucially provide an easily used method to identify 
whether these two information sources are in conflict in a
way which is of consequence to the whole system reliability
estimate.

These methodological contributions can be immediately used in
applications by reliability practitioners as 
we provide easy to use software tools.

An important next step is to extend the model to include right-censored observations which are common in the reliability setting.
In particular, this allows to use component failure observations from a running system
to calculate its remaining useful life.
We see two potential approaches.
First, to obtain lower and upper system reliability bounds 
one can assume that a component either fails immediately after censoring or
continues to function during the entire time horizon.
This minimal assumption will be simple to implement but will lead to high imprecision.
Alternatively, one can assume exchangeability with other surviving components at the moment of censoring.
This approach will be more complex to accomodate but will lead to less imprecision.
Indeed, this assumption lies at the core of the Kaplan-Meier estimator \citep{1958:kaplan-meier},
and has already been adopted by \citet{2004:coolen-yan} in an imprecise probability context.

Upscaling the survival signature to large real-world systems and networks, consisting of thousands of components, is a major challenge.
However, even for such systems the fact that one only needs to derive the survival signature once for a system is an advantage,
and also the monotonicity of the survival signature for coherent systems is very useful if one can only derive it partially.  

The survival signature and its use for uncertainty quantification for system
reliability can be generalized quite straightforwardly, mainly due to the simplicity of this concept.
For example, one may generalize the system structure function from a binary function to a probability, 
to reflect uncertainty about system functioning for known states of its components, with a further
generalization to imprecise probabilities possible.

\section*{Acknowledgements}

Gero Walter was supported by the DINALOG project
``Coordinated Advanced Maintenance and Logistics Planning for the Process Industries'' (CAMPI).

Louis Aslett was supported by the i-like project (EPSRC grant reference number EP/K014463/1).


\section*{References}

\bibliographystyle{elsarticle-harv}
\bibliography{ijar-npb-refs}

\begin{thebibliography}{22}
\expandafter\ifx\csname natexlab\endcsname\relax\def\natexlab#1{#1}\fi
\expandafter\ifx\csname url\endcsname\relax
  \def\url#1{\texttt{#1}}\fi
\expandafter\ifx\csname urlprefix\endcsname\relax\def\urlprefix{URL }\fi

\bibitem[{Aslett(2016)}]{2015:aslett-RT}
Aslett, L., 2016. {ReliabilityTheory}: Tools for structural reliability
  analysis. R package.
\newline\urlprefix\url{http://www.louisaslett.com}

\bibitem[{Aslett et~al.(2015)Aslett, Coolen, and Wilson}]{2015:bayessurvsign}
Aslett, L., Coolen, F., Wilson, S., 2015. Bayesian inference for reliability of
  systems and networks using the survival signature. Risk Analysis 35,
  1640--1651.
\newline\urlprefix\url{http://dx.doi.org/10.1111/risa.12228}

\bibitem[{Augustin et~al.(2014)Augustin, Coolen, de~Cooman, and
  Troffaes}]{itip}
Augustin, T., Coolen, F., de~Cooman, G., Troffaes, M., 2014. Introduction to
  Imprecise Probabilities. Wiley, New York.

\bibitem[{Barlow and Proschan(1975)}]{BP75}
Barlow, R., Proschan, F., 1975. Statistical Theory of Reliability and Life
  Testing. Holt, Rinehart and Winston, Inc., New York.

\bibitem[{Bernardo and Smith(2000)}]{2000:bernardosmith}
Bernardo, J., Smith, A., 2000. Bayesian Theory. Wiley, Chichester.

\bibitem[{Bickel(2015)}]{2015:bickel}
Bickel, D., 2015. Inference after checking multiple {B}ayesian models for data
  conflict and applications to mitigating the influence of rejected priors.
  International Journal of Approximate Reasoning 66, 53--72.
\newline\urlprefix\url{http://dx.doi.org/10.1016/j.ijar.2015.07.012}

\bibitem[{Coolen(2011)}]{Co11}
Coolen, F., 2011. Nonparametric predictive inference. In: Lovric, M. (Ed.),
  International Encyclopedia of Statistical Science. Springer, Berlin, pp.
  968--970.

\bibitem[{Coolen and Coolen-Maturi(2012)}]{2012:survsign}
Coolen, F., Coolen-Maturi, T., 2012. Generalizing the signature to systems with
  multiple types of components. In: Zamojski, W., Mazurkiewicz, J., Sugier, J.,
  Walkowiak, T., Kacprzyk, J. (Eds.), Complex Systems and Dependability. Vol.
  170 of Advances in Intelligent and Soft Computing. Springer, pp. 115--130.

\bibitem[{Coolen and Coolen-Maturi(2015)}]{CCM15}
Coolen, F., Coolen-Maturi, T., 2015. Predictive inference for system
  reliability after common-cause component failures. Reliability Engineering
  and System Safety 135, 27--33.

\bibitem[{Coolen et~al.(2014)Coolen, Coolen-Maturi, and Al-nefaiee}]{CCMA14}
Coolen, F., Coolen-Maturi, T., Al-nefaiee, A., 2014. Nonparametric predictive
  inference for system reliability using the survival signature. Journal of
  Risk and Reliability 228, 437--448.

\bibitem[{Coolen and Yan(2004)}]{2004:coolen-yan}
Coolen, F., Yan, K., 2004. Nonparametric predictive inference with
  right-censored data. Journal of Statistical Planning and Inference 126,
  25--54.

\bibitem[{De~Finetti(1974)}]{DF74}
De~Finetti, B., 1974. Theory of Probability. Wiley, Chichester.

\bibitem[{Evans and Moshonov(2006)}]{2006:evans}
Evans, M., Moshonov, H., 2006. Checking for prior-data conflict. Bayesian
  Analysis 1, 893--914.
\newline\urlprefix\url{http://projecteuclid.org/euclid.ba/1340370946}

\bibitem[{Kaplan and Meier(1958)}]{1958:kaplan-meier}
Kaplan, E., Meier, P., 1958. Nonparametric estimation from incomplete
  observations. Journal of the American Statistical Association 53, 457--481.

\bibitem[{Klenke and Mattner(2010)}]{klenke2010}
Klenke, A., Mattner, L., 2010. Stochastic ordering of classical discrete
  distributions. Advances in Applied Probability 42~(2), 392--410.

\bibitem[{{R Core Team}(2016)}]{R}
{R Core Team}, 2016. R: A Language and Environment for Statistical Computing. R
  Foundation for Statistical Computing, Vienna, Austria.
\newline\urlprefix\url{https://www.R-project.org/}

\bibitem[{Samaniego(2007)}]{Sa07}
Samaniego, F., 2007. System Signatures and their Applications in Engineering
  Reliability. Springer, New York.

\bibitem[{Shaked and Shanthikumar(2007)}]{shaked2007}
Shaked, M., Shanthikumar, J., 2007. Stochastic orders, 1st Edition. Springer,
  New York.

\bibitem[{Troffaes et~al.(2013)Troffaes, Walter, and
  Kelly}]{Troffaes2013a-short}
Troffaes, M., Walter, G., Kelly, D., 2013. A robust {B}ayesian approach to
  modelling epistemic uncertainty in common-cause failure models. Reliability
  Engineering \& System Safety 125, 13--21.
\newline\urlprefix\url{http://dx.doi.org/10.1016/j.ress.2013.05.022}

\bibitem[{Walley(1991)}]{1991:walley}
Walley, P., 1991. Statistical Reasoning with Imprecise Probabilities. Chapman
  and Hall, London.

\bibitem[{Walter(2013)}]{2013:diss-gw}
Walter, G., 2013. Generalized {B}ayesian inference under prior-data conflict.
  Ph.D. thesis, Department of Statistics, LMU Munich.
\newline\urlprefix\url{http://edoc.ub.uni-muenchen.de/17059/}

\bibitem[{Walter and Augustin(2009)}]{2009:WalterAugustin}
Walter, G., Augustin, T., 2009. Imprecision and prior-data conflict in
  generalized {B}ayesian inference. Journal of Statistical Theory and Practice
  3, 255--271.
\newline\urlprefix\url{http://dx.doi.org/10.1080/15598608.2009.10411924}

\end{thebibliography}

\section*{Appendix}
\appendix
\renewcommand*{\thesection}{\Alph{section}}

\section{Proofs}
\label{ap:proofs}

\begin{proof}[\textbf{Proof of Theorem \ref{thm:y}, p\pageref{thm:y}}]
  \label{prf:y}
  Consider the likelihood ratio for the two Beta Binomial distributions $\beta_{\yu}$ and $\beta_{\yl}$,
  \begin{align*}
    \lefteqn{\mathcal{L}(l) := \frac{p(l \mid \yu, n, m, s, N)}{p(l \mid \yl, n, m, s, N)}} \\
    &= \frac{B(l + n \yu + s, m - l + n (1 - \yu) + N - s) B(n \yl + s, n (1 - \yl) + N - s)}
            {B(n \yu + s, n (1 - \yu) + N - s) B(l + n \yl + s, m - l + n (1 - \yl) + N - s)} \\
    &= \frac{\Gamma(l + n \yu + s) \Gamma(m - l + n (1 - \yu) + N - s) \Gamma(n \yl + s) \Gamma(n (1 - \yl) + N - s)}
            {\Gamma(l + n \yl + s) \Gamma(m - l + n (1 - \yl) + N - s) \Gamma(n \yu + s) \Gamma(n (1 - \yu) + N - s)} \\
    &= \left\{ \begin{aligned}
         \frac{\prod_{x=0}^{m-1} (x + n (1 - \yu) + N - s)}
              {\prod_{x=0}^{m-1} (x + n (1 - \yl) + N - s)} &\quad\mbox{ for } l=0 \\
         \frac{\prod_{x=0}^{l-1} (x + n \yu + s) \prod_{x=0}^{m-l-1} (x + n (1 - \yu) + N - s)}
              {\prod_{x=0}^{l-1} (x + n \yl + s) \prod_{x=0}^{m-l-1} (x + n (1 - \yl) + N - s)} &\quad\mbox{ for } 0<l<m \\
         \frac{\prod_{x=0}^{m-1} (x + n \yu + s)}
              {\prod_{x=0}^{m-1} (x + n \yl + s)} &\quad\mbox{ for } l=m
       \end{aligned} \right.
  \end{align*}
  since $\Gamma(x+1)=x \Gamma(x)$.
  
  Thus,
  \begin{align*}
    \frac{\mathcal{L}(l+1)}{\mathcal{L}(l)} &=
      \frac{(l + n \yu + s) (m - l - 1 + n (1 - \yl) + N - s)}
           {(l + n \yl + s) (m - l - 1 + n (1 - \yu) + N - s)} \\
    &> 1 \quad\mbox{when}\quad 0 \le \yl < \yu \le 1
  \end{align*}
    
  Hence, $\mathcal{L}(\cdot)$ is monotone increasing for $0 < \yl < \yu < 1$, so that $\beta_{\yu}$ is larger than or equal to $\beta_{\yl}$ in monotone likelihood ratio order ($\beta_{\yu} \ge_\mathrm{lr} \beta_{\yl}$).  But, $\beta_{\yu} \ge_\mathrm{lr} \beta_{\yl} \implies \beta_{\yu} \ge_\mathrm{st} \beta_{\yl}$ (\cite[Theorem 1.C.1, p.43]{shaked2007}) giving the required result.
\end{proof}


\begin{proof}[\textbf{Proof of Theorem \ref{thm:n}, p\pageref{thm:n}}]
  \label{prf:n}
  Consider the likelihood ratio for the two Beta Binomial distributions $\beta_{\nu}$ and $\beta_{\nl}$,
  \begin{align*}
    \lefteqn{\mathcal{L}(l) := \frac{p(l \mid y, \nu, m, s, N)}{p(l \mid y, \nl, m, s, N)}} \\
    &= \frac{B(l + \nu y + s, m - l + \nu (1 - y) + N - s) B(\nl y + s, \nl (1 - y) + N - s)}
            {B(\nu y + s, \nu (1 - y) + N - s) B(l + \nl y + s, m - l + \nl (1 - y) + N - s)} \\
    &= \frac{\Gamma(l + \nu y + s) \Gamma(m - l + \nu (1 - y) + N - s)}
            {\Gamma(l + \nl y + s) \Gamma(m - l + \nl (1 - y) + N - s)} \\
    &\quad \times \frac{\Gamma(\nl y + s) \Gamma(\nl (1 - y) + N - s) \Gamma(\nu+N) \Gamma(m+\nl+N)}
                       {\Gamma(\nu y + s) \Gamma(\nu (1 - y) + N - s)\Gamma(\nl+N) \Gamma(m+\nu+N)} \\
    &= \left\{ \begin{aligned}
         \frac{\prod_{x=0}^{m-1} (x + \nu (1 - y) + N - s) \prod_{x=0}^{m-1} (x + \nl + N)}
              {\prod_{x=0}^{m-1} (x + \nl (1 - y) + N - s) \prod_{x=0}^{m-1} (x + \nu + N)} &\quad\mbox{ for } l=0 \\
         \frac{\prod_{x=0}^{l-1} (x + \nu y + s) \prod_{x=0}^{m-l-1} (x + \nu (1 - y) + N - s)}
              {\prod_{x=0}^{l-1} (x + \nl y + s) \prod_{x=0}^{m-l-1} (x + \nl (1 - y) + N - s)} \\
         \quad \times \frac{\prod_{x=0}^{m-1} (x + \nl + N)}{\prod_{x=0}^{m-1} (x + \nu + N)} &\quad\mbox{ for } 0<l<m \\
         \frac{\prod_{x=0}^{m-1} (x + \nu y + s) \prod_{x=0}^{m-1} (x + \nl + N)}
              {\prod_{x=0}^{m-1} (x + \nl y + s) \prod_{x=0}^{m-1} (x + \nu + N)} &\quad\mbox{ for } l=m
       \end{aligned} \right.
  \end{align*}
  since $\Gamma(x+1)=x \Gamma(x)$.
  
  Thus,
  \[ \frac{\mathcal{L}(l+1)}{\mathcal{L}(l)} =
      \frac{(l + \nu y + s) (m - l - 1 + \nl (1 - y) + N - s)}
           {(l + \nl y + s) (m - l - 1 + \nu (1 - y) + N - s)}
  \]
  
  However, unlike the case for the $y$ parameter in Theorem \ref{thm:y}, neither $\beta_{\nl}$ 
  nor $\beta_{\nu}$ can be guaranteed to dominate for all possible values 
  for the other parameters, so that necessary conditions for
  monotonicity (either increasing or decreasing) must be established.
  We require,
  \[ \frac{(l + \nu y + s) (m - l - 1 + \nl (1 - y) + N - s)}
          {(l + \nl y + s) (m - l - 1 + \nu (1 - y) + N - s)}
     >1
  \]
  After extensive routine algebra, this can be conveniently expressed as
  \begin{align*}
    (\nu - \nl)[y (N + m - 1) - s] - l (\nu - \nl)>0.
  \end{align*}
  This limit is hardest to satisfy for $l=m-1$ since $\nu-\nl>0$ (note $l\ne m$ since we are evaluating $\mathcal{L}(l+1)/\mathcal{L}(l)$, so $m-1$ is the maximal value $l$ can take).
  Thus, for monotonicity to hold for all $l$, we require
  \begin{align*}
    (\nu - \nl)[y (N + m - 1) - s] - (m - 1) (\nu - \nl) &>0 \\
    \implies (\nu - \nl)[y (N + m - 1) - s - m + 1] &>0
  \end{align*}
  Since $\nu-\nl>0$ by definition, we have a monotonically increasing 
  likelihood ratio only when \[ y (N + m - 1) - s - m + 1 > 0. \]
  By a similar argument, the likelihood ratio is only monotonically decreasing when
  \[ y (N + m - 1) - s < 0. \]
  Thus,
  \begin{equation}
    y > \frac{s + m - 1}{N + m - 1} \implies \beta_{\nu} \ge_{\mathrm{lr}} \beta_{\nl} \implies \beta_{\nu} \ge_{\mathrm{st}} \beta_{\nl}
    \label{eq:nuDOMnl}
  \end{equation}
  and 
  \begin{equation}
    y < \frac{s}{N + m - 1} \implies \beta_{\nu} \le_{\mathrm{lr}} \beta_{\nl} \implies \beta_{\nu} \le_{\mathrm{st}} \beta_{\nl} 
    \label{eq:nlDOMnu}
  \end{equation}
  by \cite[Theorem 1.C.1, p.43]{shaked2007}.  In the intermediate case,
  \[ \frac{s}{N+m-1} < y < \frac{s+m-1}{N+m-1} \]
  standard likelihood ratio ordering theory cannot definitively state
  the stochastic ordering on $\beta_{\nu}$ and $\beta_{\nl}$.
\end{proof}

\begin{proof}[\textbf{Proof of Lemma \ref{lem:n}, p\pageref{lem:n}}]
  \label{prf:n2}
  \eqref{eq:nuDOMnl} and \eqref{eq:nlDOMnu}
  are sufficient but not necessary conditions.  Using theory in 
  \cite{klenke2010} we can sharpen these conditions to provide
  first-order stochastic dominance conditions for a larger range of
  parameter values.
  
  Proposition 2.1, p.399 of \cite{klenke2010} proves that half-monotone
  likelihood ratio ordering --- i.e.\ monotonicity of $\mathcal{L}(l+1) / \mathcal{L}(l)$ ---
  together with left and right tail conditions on $\mathcal{L}(\cdot)$
  imply first order stochastic dominance.
  
  \medskip
  \noindent{\emph{Half-monotonicity of $\mathcal{L}(\cdot)$}}
  
  Although there exist parameters for which $\mathcal{L}(\cdot)$ is
  not monotone, it is half-monotone.  That is, $\mathcal{L}(l+1) / \mathcal{L}(l)$ is itself monotone.  For simplicity, write
  \begin{align*}
    \frac{\mathcal{L}(l+1)}{\mathcal{L}(l)} = \frac{(l + \pu)(\el - l)}{(l + \pl)(\eu - l)}
    \quad \mbox{where} \left\{\begin{aligned}
      \pu &= \nu y + s \\
      \pl &= \nl y + s \\
      \eu &= m - 1 + \nu (1 - y) + N - s \\
      \el &= m - 1 + \nl (1 - y) + N - s
    \end{aligned} \right.
  \end{align*}
  Then,
  \begin{align*}
    \frac{\mathcal{L}(l+2)/\mathcal{L}(l+1)}{\mathcal{L}(l+1)/\mathcal{L}(l)} =
    \frac{(\pl + l)(\pu + l + 1)(\el - l - 1)(\eu - l)}
           {(\pl + l + 1)(\pu + l)(\el -
l)(\eu - l - 1)}
    &< 1 \\
    \iff \qquad \frac{\pl + l}{\pu + l} \cdot \frac{\el - l - 1}{\eu - l - 1} 
    &< \frac{\pl + l + 1}{\pu + l + 1} \cdot \frac{\el -
l}{\eu - l}
  \end{align*}
  But, $\pu > \pl > 0$, $\eu > \el > 0$, $l > 0$, 
  so it is trivial to prove
  \[
    \frac{\pl + l}{\pu + l} < \frac{\pl + l + 1}{\pu + l + 1} \qquad\mbox{and}\qquad \frac{\el - l - 1}{\eu - l - 1} < \frac{\el -
l}{\eu - l} \qquad \forall\ l \in \{0,\dots,m\}
  \]
  Thus we can conclude that $\mathcal{L}(\cdot)$ is half monotone decreasing.

  \medskip
  \noindent{\emph{Tail conditions on $\mathcal{L}(\cdot)$}}
  
  It is not difficult to derive the same loose bounds as in Theorem \ref{thm:n}
  using the tail conditions.  However, it is also easy to see that 
  these are sufficient but not necessary.  Sharpening these bounds
  in terms of the other parameter values involves seemingly intractable
  algebra, so we leave the tail condition as the alternative slightly 
  more costly numerical check when the conditions of Theorem
  \ref{thm:n} are not satisfied.  Evaluation of $\mathcal{L}(\cdot)$ 
  at two values is still orders of magnitude less costly than 
  reevaluation of $\lRsys(\cdot)$ or $\uRsys(\cdot)$.
\end{proof}

\section{Software details}
\label{ap:software}

Functions which make it easy to use the methods of this paper
have been added to the \textbf{R} package \texttt{ReliabilityTheory} \citep{2015:aslett-RT}.
There are two functions of particular note: \texttt{computeSystemSurvivalSignature}
and, implementing the result from Appendix~\ref{ap:proofs} above, \texttt{nonParBayesSystemInferencePriorSets}.

\subsection{Computing the survival signature}

The function \texttt{computeSystemSurvivalSignature} allows easy 
computation of the survival signature if the system is expressed as an
undirected graph with `start' and `terminal' nodes (which are not
considered components for survival signature computation).  The system
is considered to work if there is a path from the start to the terminal
node passing only through functioning components.

Graph representations of systems are most simply defined by using the 
\texttt{graph.formula} function.  The `start' node should be denoted
\texttt{s} and the `terminal' node should be denoted \texttt{t} and
intermediate nodes (representing actual components) should be numbered and
connected by edges denoted by \texttt{-}, where the numbering denotes physically
distinct components.  Component numbers can be repeated to include
multiple links.  For example, to build a simple three component series
system:

\noindent\texttt{sys <- graph.formula(s\,-\,1\,-\,2\,-\,3\,-\,t)}

and to build a three component parallel system:

\noindent\texttt{sys <- graph.formula(s\,-\,1\,-\,t, s\,-\,2\,-\,t, s\,-\,3\,-\,t)}

There is an additional shorthand which indicates a link exists to a list of multiple components separated by the \texttt{:} operator, so that the parallel system can be also be expressed more compactly by:

\noindent\texttt{sys <- graph.formula(s\,-\,1:2:3\,-\,t)}

Therefore, the simple bridge system of Figure \ref{fig:bridge-layout} can be constructed with:

\noindent\texttt{sys <- graph.formula(s\,-\,1\,-\,2\,-\,3\,-\,t, s\,-\,4\,-\,5\,-\,3\,-\,t, 1:4\,-\,6\,-\,2:5)}
\begin{itemize}
  \item \texttt{s\,-\,1\,-\,2\,-\,3\,-\,t} signifies the route from left to right entering the first component going across the top of the system block diagram in Figure \ref{fig:bridge-layout};
  \item \texttt{s\,-\,4\,-\,5\,-\,3\,-\,t} signifies the bottom route through the block diagram;
  \item \texttt{1:4\,-\,6\,-\,2:5} connects the top two components of type 3 to the bottom two components of type 3, signifying the bridge.
\end{itemize}
Naturally such as expression is not necessarily unique, so that completely equivalently one may write:

\noindent\texttt{sys <- graph.formula(s\,-\,1:4\,-\,6\,-\,2:5\,-\,3\,-\,t, 1\,-\,2, 4\,-\,5)}

\ 

With the structure defined and the individual components numbered, it 
just remains to specify the types of each component.  This can be done
using the \texttt{setCompTypes} function.  This function takes the
system graph and a list of component type names (as the tag) and corresponding
component numbers (as the value).  Thus, completing the example for Figure
\ref{fig:bridge-layout}:

\noindent\texttt{sys <- setCompTypes(sys, list("T1"=c(1,2,4,5), "T2"=c(6),}

\noindent\texttt{~~~~~~~~~~~~~~~~~~~~~~~~~~~~~~"T3"=c(3)))}

\ 

Computing the survival signature then involves a simple function call:
\noindent\texttt{survsig <- computeSystemSurvivalSignature(sys)}

\subsection{Computing sets of system survival probabilities}

Once the system has been correctly described using an undirected graph
as above, the methods presented in Sections \ref{sec:nonparamapproach}
-- \ref{sec:setsofrel} can be used via the function
\texttt{nonParBayesSystemInferencePriorSets}.

The function prototype is:

\noindent\texttt{nonParBayesSystemInferencePriorSets(at.times, survival.signature,}

\noindent\texttt{~~~~~~~~~~test.data, nLower=2, nUpper=2, yLower=0.5, yUpper=0.5)}

Aside from the system design, which can be passed to the function via
the \texttt{survival.signature} argument, the remaining elements which
must be specified are the:
\begin{enumerate}
  \item grid of times at which to evaluate the posterior, ${\cal T} = \{t_1, \ldots, \tmax\}$, via the \texttt{at.times} argument.
  \item component test data $\vec{t}^k = (t^k_1, \ldots, t^k_{n_k})$ for $k=1,\dots,K$, via the \texttt{test.data} argument.
  \item prior sets via the range of prior parameter sets $\PktZ = [\nktzl, \nktzu] \times [\yktzl, \yktzu]$, via the \texttt{nLower}, \texttt{nUpper}, \texttt{yLower} and \texttt{yUpper} arguments.
\end{enumerate}

The grid of times, \texttt{at.times}, is specified as simply a vector 
of time points.

The \texttt{test.data} argument is a list of component type names (as
the tag) and corresponding lifetime data (as the value), for example 
a toy sized dataset for each component would be expressed as:

\noindent\texttt{test.data=list("T1"=c(0.19, 0.73, 1.87, 1.17),}

\noindent\texttt{~~~~~~~~~~~~~~~"T2"=c(0.22, 0.27, 0.63, 1.80, 1.25, 1.95),}

\noindent\texttt{~~~~~~~~~~~~~~~"T3"=c(1.33, 0.65, 1.59))}

Finally, there are multiple options for specifying the prior parameter
sets.  Each of the \texttt{nLower}, \texttt{nUpper}, \texttt{yLower}
and \texttt{yUpper} arguments can be specified as:
\begin{itemize}
  \item a single value for a homogeneous prior across time and components.  e.g.\ \texttt{nLower=2} $\implies \nktzl=2 \ \forall\,k, t$
  \item a vector of values of length $|{\cal T}|$ (\texttt{length(at.times)}), for a time inhomogeneous prior which is identical across component types.
  \item a data frame of size $1\times K$, where each column is named the same as in the \texttt{survival.signature} and \texttt{test.data} arguments, for a time homogeneous prior which varies across component types.
  \item a data frame of size $|{\cal T}|\times K$, where each column is named the same as in the \texttt{survival.signature} and \texttt{test.data} arguments, for a time inhomogeneous prior which varies across component types.
\end{itemize}

With these arguments supplied,  \texttt{nonParBayesSystemInferencePriorSets} will then compute
the posterior sets automatically in parallel across the cores of a
multicore CPU and return a list with two objects, named \texttt{lower} 
and \texttt{upper}, containing respectively the lower and upper bound 
for the system reliability function $\Rsys(t)$.

\end{document}